%% file: 2022dbl.tex
\documentclass[twocolumn]{aastex631}

\usepackage{amsmath}
\usepackage{amssymb}
\usepackage{longtable}

\def\dbl{AT\,2022dbl}

\input{symbols.tex}

\begin{document}

\title{The Double Tidal Disruption Event \dbl\ Implies That at Least Some ``Standard'' Optical TDEs are Partial Disruptions}

\correspondingauthor{Lydia~Makrygianni}
\email{lydiamakr@gmail.com}

\author[0000-0002-7466-4868]{Lydia~Makrygianni}
\affiliation{The School of Physics and Astronomy, Tel Aviv University, Tel Aviv 69978, Israel}
\affiliation{Department of Physics, Lancaster University, Lancaster LA1 4YB, UK}

\author[0000-0001-7090-4898]{Iair~Arcavi}
\affiliation{The School of Physics and Astronomy, Tel Aviv University, Tel Aviv 69978, Israel}

\author[0000-0001-9570-0584]{Megan~Newsome}
\affiliation{Las Cumbres Observatory, Goleta, CA 93117, USA}
\affiliation{Department of Physics, University of California, Santa Barbara, CA 93106, USA}

\author[0000-0002-5116-844X]{Ananya~Bandopadhyay}
\affiliation{Department of Physics, Syracuse University, Syracuse, NY 13210, USA}

\author[0000-0003-3765-6401]{Eric~R.~Coughlin}
\affiliation{Department of Physics, Syracuse University, Syracuse, NY 13210, USA}

\author[0000-0002-8304-1988]{Itai~Linial}
\affiliation{Columbia Astrophysics Laboratory, Columbia University, New York, NY 10027, USA}
\affiliation{School of Natural Sciences, Institute for Advanced Study, Princeton, NJ 08540, USA}

\author[0000-0001-6350-8168]{Brenna~Mockler}
\affiliation{The Observatories of the Carnegie Institution for Science, Pasadena, CA 91101, USA}

\author[0000-0001-9185-5044]{Eliot~Quataert}
\affiliation{Department of Astrophysical Sciences, Princeton University, Peyton Hall, Princeton, NJ 08540, USA}

\author[0000-0002-2137-4146]{Chris~Nixon}
\affiliation{School of Physics and Astronomy, Sir William Henry Bragg Building, Woodhouse Ln., University of Leeds, Leeds LS2 9JT, UK}

\author[0000-0003-3766-7266]{Benjamin~Godson}
\affiliation{Department of Physics, University of Warwick, Gibbet Hill Road, Coventry, CV4 7AL, UK}

\author[0000-0003-4663-4300]{Miika~Pursiainen}
\affiliation{Department of Physics, University of Warwick, Gibbet Hill Road, Coventry, CV4 7AL, UK}

\author[0000-0002-8597-0756]{Giorgos~Leloudas}
\affiliation{DTU Space, National Space Institute, Technical University of Denmark, Elektrovej 327, 2800, Kgs. Lyngby, Denmark}

\author[0000-0002-4235-7337]{K.~Decker~French}
\affiliation{Department of Astronomy, University of Illinois, 1002 W. Green Street, Urbana, IL 61801, USA}

\author[0000-0002-0350-4488]{Adi~Zitrin}
\affiliation{Physics Department, Ben-Gurion University of the Negev, PO Box 653, Be`er-Sheva 8410501, Israel}

\author[0009-0007-8485-1281]{Sara~Faris}
\affiliation{The School of Physics and Astronomy, Tel Aviv University, Tel Aviv 69978, Israel}

\author[0000-0002-9347-2298]{Marco~C.~Lam}
\affiliation{Institute for Astronomy, University of Edinburgh, Royal Observa- tory, Blackford Hill, Edinburgh EH9 3HJ, UK}

\author[0000-0002-5936-1156]{Assaf~Horesh}
\affiliation{Racah Institute of Physics, The Hebrew University of Jerusalem, Jerusalem 91904, Israel}

\author[0000-0003-0466-3779]{Itai~Sfaradi}
\affiliation{Racah Institute of Physics, The Hebrew University of Jerusalem, Jerusalem 91904, Israel}

\author[0000-0002-9113-7162]{Michael~Fausnaugh}
\affiliation{Department of Physics \& Astronomy, Texas Tech University, Lubbock, TX 79409-1051, USA}

\author[0000-0002-4534-7089]{Ehud~Nakar}
\affiliation{The School of Physics and Astronomy, Tel Aviv University, Tel Aviv 69978, Israel}

\author[0000-0002-8648-0767]{Kendall~Ackley}
\affiliation{Department of Physics, University of Warwick, Gibbet Hill Road, Coventry, CV4 7AL, UK}

\author[0000-0002-1895-6639]{Moira~Andrews}
\affiliation{Las Cumbres Observatory, Goleta, CA 93117, USA}
\affiliation{Department of Physics, University of California, Santa Barbara, CA 93106, USA}

\author[0000-0002-0326-6715]{Panos~Charalampopoulos}
\affiliation{Department of Physics and Astronomy, University of Turku, FI-20014 Turku, Finland}

\author[0009-0000-5659-9006]{Benjamin~D.~R.~Davies}
\affiliation{Department of Physics, University of Warwick, Gibbet Hill Road, Coventry, CV4 7AL, UK}
\affiliation{center for Exoplanets and Habitability, University of Warwick, Gibbet Hill Road, Coventry CV4 7AL, UK}

\author[0000-0002-7579-1105]{Yael~Dgany}
\affiliation{The School of Physics and Astronomy, Tel Aviv University, Tel Aviv 69978, Israel}

\author[0000-0003-3665-5482]{Martin~J.~Dyer}
\affiliation{Department of Physics and Astronomy, University of Sheffield, Sheffield, S3 7RH, UK}

\author[0000-0003-4914-5625]{Joseph~Farah}
\affiliation{Las Cumbres Observatory, Goleta, CA 93117, USA}
\affiliation{Department of Physics, University of California, Santa Barbara, CA 93106, USA}

\author{Rob~Fender}
\affiliation{Astrophysics, Department of Physics, University of Oxford, Keble Road, Oxford OX1 3RH, UK}

\author[0000-0003-3189-9998]{David~A.~Green}
\affiliation{Astrophysics Group, Cavendish Laboratory, 19 J.J. Thomson Avenue, Cambridge CB3 0HE, UK}

\author[0000-0003-4253-656X]{D.~Andrew~Howell}
\affiliation{Las Cumbres Observatory, Goleta, CA 93117, USA}
\affiliation{Department of Physics, University of California, Santa Barbara, CA 93106, USA}

\author[0000-0002-0440-9597]{Thomas~Killestein}
\affiliation{Department of Physics and Astronomy, University of Turku, FI-20014 Turku, Finland}

\author[0009-0007-7151-7313]{Niilo~Koivisto}
\affiliation{Department of Physics and Astronomy, University of Turku, FI-20014 Turku, Finland}

\author[0000-0002-3464-0642]{Joseph~Lyman}
\affiliation{Department of Physics, University of Warwick, Gibbet Hill Road, Coventry, CV4 7AL, UK}

\author[0000-0001-5807-7893]{Curtis~McCully}
\affiliation{Las Cumbres Observatory, Goleta, CA 93117, USA}

\author[0009-0004-6130-7775]{Morgan~A.~Mitchell}
\affiliation{Department of Physics, University of Warwick, Gibbet Hill Road, Coventry, CV4 7AL, UK}
\affiliation{center for Exoplanets and Habitability, University of Warwick, Gibbet Hill Road, Coventry CV4 7AL, UK}

\author[0000-0003-0209-9246]{Estefania~Padilla~Gonzalez}
\affiliation{Las Cumbres Observatory, Goleta, CA 93117, USA}
\affiliation{Department of Physics, University of California, Santa Barbara, CA 93106, USA}

\author{Lauren~Rhodes}
\affiliation{Astrophysics, Department of Physics, University of Oxford, Keble Road, Oxford OX1 3RH, UK}

\author{Anwesha~Sahu}
\affiliation{Department of Physics, University of Warwick, Gibbet Hill Road, Coventry, CV4 7AL, UK}

\author[0000-0003-0794-5982]{Giacomo~Terreran}
\affiliation{Las Cumbres Observatory, Goleta, CA 93117, USA}
\affiliation{Department of Physics, University of California, Santa Barbara, CA 93106, USA}

\author[0009-0005-8379-3871s s]{Ben~Warwick}
\affiliation{Department of Physics, University of Warwick, Gibbet Hill Road, Coventry, CV4 7AL, UK}

\begin{abstract}

Flares produced following the tidal disruption of stars by supermassive black holes can reveal the properties of the otherwise dormant majority of black holes and the physics of accretion. In the past decade, a class of optical-ultraviolet tidal disruption flares has been discovered whose emission properties do not match theoretical predictions. This has led to extensive efforts to model the dynamics and emission mechanisms of optical-ultraviolet tidal disruptions in order to establish them as probes of supermassive black holes. Here we present the optical-ultraviolet tidal disruption event \dbl, which showed a nearly identical repetition 700 days after the first flare. Ruling out gravitational lensing and two chance unrelated disruptions, we conclude that at least the first flare represents the partial disruption of a star, possibly captured through the Hills mechanism. Since both flares are typical of the optical-ultraviolet class of tidal disruptions in terms of their radiated energy, temperature, luminosity, and spectral features, it follows that either the entire class are partial rather than full stellar disruptions, contrary to the prevalent assumption, or that some members of the class are partial disruptions, having nearly the same observational characteristics as full disruptions. Whichever option is true, these findings could require revised models for the emission mechanisms of optical-ultraviolet tidal disruption flares and a reassessment of their expected rates.

\end{abstract}


\section{Introduction} \label{sec:intro}

After a star is torn apart by the tidal forces of a supermassive black hole (SMBH), in what is known as a Tidal Disruption Event (TDE), typically half of its mass will be bound to the black hole and half will be ejected \citep{lacy82, Rees1988}. For SMBH masses $\lesssim10^8\Msun$ (with the exact threshold depending on the spin of the SMBH), the disruption of a Sun-like star will occur outside the event horizon, producing an observable flare. Such flares can reveal the properties of the otherwise dormant majority of black holes and the physics of accretion (see \citealt{Gezari2021_rev} for a recent review of TDEs).

If the star is completely disrupted, the rate at which the bound material falls onto the black hole scales as $t^{-5/3}$ at late times \citep{Rees1988,Phinney1989}. For partial disruptions the fallback rate can be steeper \citep{GR2013,CN2019,Goicovic2019,Ryu2020a} reaching $t^{-9/4}$ at late times \citep{CN2019}. This material is expected to form an accretion disk that emits X-ray radiation \citep{cannizzo90}, as was indeed observed (see \citealt{Saxton2021SSRv} for a recent review of X-ray TDEs). 

In the past decade, however, another class of TDEs has been found that, surprisingly, emit mostly in the optical and ultraviolet (\citealt{Gezari2012,Arcavi2014}; see \citealt{VV2020} for a recent review of optical-ultraviolet TDEs). This class of events exhibit roughly constant blackbody temperatures of a few $\times\,10^4$\,K (1--2 orders of magnitude smaller than expected from an accretion disk), blackbody radii of $\sim10^{15}$\,cm (two orders of magnitude larger than the tidal disruption radius of a Sun-like star), total released energies of $10^{50}$--$10^{51}$\,erg (2--3 orders of magnitude lower than the expected energy released by the accretion of a solar mass of material at 10\% efficiency; the so-called ``missing energy problem'') and broad H and/or He emission features in their spectra (which were not predicted). Yet, these transients occur in otherwise quiescent and non-starforming galaxies, disfavoring extreme active galactic nuclei (AGN) variability and massive-star explosions as their origin. More strikingly, their rates drop dramatically (faster than the SMBH mass function) in galaxies hosting SMBHs with masses above $10^8\Msun$ \citep{vanVelzen2018,Yao2023} providing ``smoking gun'' evidence that these transients ``know'' about the SMBH event horizon. Thus, both analytical models and numerical simulations of TDEs have tried to reconcile the discrepancies between TDE theory and observations of optical-ultraviolet TDEs either by invoking material to reprocess the emission from an accretion disk \citep{Guillochon2014,Roth2016,Dai2018,Mockler2019} or by associating the emission with the collision of stellar debris streams before the accretion disk is formed \citep{Piran2015}.

Adding to the puzzle, a few optical transients, all in galaxy centers, were recently suggested to be repeating TDEs. ASASSN-14ko \citep{Payne2021} shows tens of flares with a period of approximately 114 days which slowly decreases with time \citep[this period decrease has been explained using energy imparted to the surviving stellar core in the form of rotation, which should stop once the core is roughly ``tidally locked'' at pericenter;][]{Bandopadhyay2024}. The optical spectra of ASASSN-14ko are not typical of optical-ultraviolet TDEs, but are more similar to those of active galactic nuclei. An alternative explanation has been proposed, in which ASASSN-14ko is due to a star punching through an existing accretion disk around a black hole, with the period decrease due to hydrodynamical drag \citep{Linial_2024a}.

AT\,2019aalc \citep{Veres2024} and AT\,2021aeuk \citep{Sun2025} show double  flares within 3--4 years, but are both in previously known AGN and show spectra similar to AGN and Bowen Fluorescence Flares \citep[BFFs;][]{Trakhtenbrot2019a}. The nature of BFFs is not yet clear, but their occurrence in AGN \citep{Makrygianni2023} suggests they could be related to accretion disk instabilities rather than TDEs. 

AT\,2018fyk \citep{Wevers2019,Wevers2023} was seen to re-brighten in the X-rays and ultraviolet wavelengths approximately 1200 days after its first flare \citep{Wevers2023}. However, the first flare of AT\,2018fyk was atypical of optical-ultraviolet TDEs, having a double peak structure in its light curve and Fe lines in its spectra. This event has also been suggested as a possible single TDE around a SMBH binary \citep{Wen2024}.

AT\,2020vdq \citep{Somalwar2023} showed two flares separated by 2.6 years, discovered through a dedicated search for repeating TDEs, with the second flare showing spectra typical of optical-ultraviolet TDEs \citep{Somalwar2023}. Unfortunately, there are no published optical spectra taken during the first flare. It has therefore not been determined whether the first flare of AT\,2020vdq was due to an unrelated transient. Type Ia supernovae, for example, are $\sim$50 times more common observationally than TDEs in galaxy centers \citep{Dgany2023}, and one was indeed observed within 2 years of the TDE AT\,2021mhg at the same position \citep{Somalwar2023}. In addition, for the first flare of AT\,2020vdq to be a TDE, it would need be one of the faintest and lowest-temperature TDEs ever observed \citep{Somalwar2023}. Even if both flares of AT\,2020vdq were indeed TDEs, it is not trivial to associate them to the disruption of the same star, given that the two flares differ substantially in their photometric properties, and that their spectral similarity can not be determined. In addition, the host galaxy of AT\,2020vdq is a post-starburst galaxy. Such galaxies have been shown to exhibit an elevated TDE rate \citep{Arcavi2014,French2016,French2020SSRv}, which could produce two unrelated flares on the observed timescales at a non-negligible probability (as we show here in Section \ref{sec:two_unrelated_tdes}). 

\dbl\ \citep{Stanek2022,Arcavi2022} is an otherwise ``standard'' optical-ultraviolet TDE with a nearly identical repetition 700 days after the first flare \citep{TNSF22024}. Both of the flares are very similar photometrically, and nearly identical spectroscopically, such that \cite{Lin2024} and \cite{Hinkle2024} claim (after ruling out an AGN origin) that both flares are due to the repeated disruption of the same star. However, the similarity between the flares could in principle be driven by the SMBH, with the two disruptions being of unrelated stars. In addition, the host of \dbl\ is a quiescent Balmer-strong galaxy. Such galaxies also show an elevated TDE rate. Albeit lower than the TDE rate enhancement seen in post-starburst galaxies \citep{French2016,French2020SSRv}, this specific galaxy was pre-selected as a potential TDE host due to its quiescent Balmer-strong properties \citep{French2018}. It remains to be shown how likely it is to observe two unrelated flares from such a galaxy on the observed timescales in order to securely determine the nature of \dbl.

Here, we present new observations and analysis of \dbl\ and new analysis of its host galaxy, as well as of the host galaxy of AT\,2020vdq. Our results securely establish \dbl\ as the first robust case of a repeating ``normal'' optical-ultraviolet TDE. We further discuss the far-reaching implications of this conclusion on the entire class of optical-ultraviolet TDEs.

\section{Discovery and Classification}

On 2022 February 17 (UT used throughout) we identified ZTF18aabdajx in the Zwicky Transient Facility \citep[ZTF;][]{Bellm2019} alert stream as a potential TDE, after a significant brightening detected on 2022 February 13 (MJD = 59623.37), using search criteria tailored to find TDEs \citep{Dgany2023}. The event is located at the center of the $z=0.0284$ galaxy WISEA J122045.05+493304.7, which, as mentioned above, was pre-selected as a potential TDE host due to its quiescent Balmer-strong properties \citep{French2018}, similar to those of other TDE host galaxies  \citep{Arcavi2014,French2016,French2020SSRv}. We consider the two $r$-band detections at the same position by ZTF in 2018, which were reported to the Transient Name Server (TNS) and received the name AT\,2018mac, as due to image subtraction artifacts. \dbl\ was spectroscopically classified on 2022 February 21 as a member of the optical-ultraviolet TDE class \citep{TNS2022NOTE}. On 2022 February 22 the event was reported \citep{Stanek2022} to the TNS following an independent detection by the All-Sky Automated Survey for Supernovae \citep[ASAS-SN;][]{Shappee2014} and it received the name \dbl. A second flare of \dbl\ was reported to the TNS \citep{TNSF22024} on 2024 February 6.

\section{Observations}

\subsection{Optical, Ultraviolet and Mid-Infrared Photometry}

We obtained optical photometry in the $BVgri$ bands, starting on MJD 59632 with the our own program on the Las Cumbres Observatory global network of 1\,m telescopes \citep{Brown2013}. We obtained reference images from Las Cumbres Observatory on MJD 59951, approximately 1 year after the first flare was detected. After the detection of the second flare, we re-started the follow-up of the event on MJD 60473. We used the \textsc{lcogtsnpipe} image subtraction pipeline \citep{Valenti2016} in order to remove host-galaxy emission. The pipeline generates the Point Spread Function (PSF) for each image and uses an implementation of the High Order Transform of PSF ANd Template Subtraction \citep[\textsc{hotpants};][]{Becker2015} to perform image subtraction and then PSF photometry at the source position. $B$ and $V$-band photometry were calibrated to the Vega system using the American Association of Variable Star Observers (AAVSO) Photometric All-Sky Survey (APASS) catalogue \citep{APASS2016}, while $gri$-band photometry was calibrated to the AB system using the SDSS Data Release (DR) 14 catalogue \citep{SDSSDR14}. 

We further retrieved PSF-fit photometry of reference-subtracted images from the ZTF forced photometry service between MJD 58194 and 60422, and from the Asteroid Terrestrial-impact Last Alert System \citep[ATLAS;][]{Tonry2018} forced photometry service between MJD 57232 and 60438 at the position of \dbl. The last pre-discovery non-detection upper limits at the position of \dbl\ were obtained by ZTF approximately 100 days before the first detection with 3$\sigma$ non-detection limits of 20.8 mag in the $r$ band and 20.7 mag in the $g$ band. 

Target-of-opportunity observations (PIs Arcavi, Makrygianni, Jiang, Hammerstein and Lin) were obtained with the Neil Gehrels Swift Observatory \citep[hereafter, \swift;][]{Gehrels2004} Optical/Ultraviolet Telescope \citep[UVOT;][]{Roming2005}. In total 81 epochs of imaging were obtained, spanning from MJD 59637 to 60470, in all six UVOT filters ($UVW2$, $UVM2$, $UVW1$, $U$, $B$ and $V$). We extracted photometry from the UVOT images using the HEAsoft software package \citep{heasoft} and the standard analysis task \texttt{uvotsource}. We performed aperture photometry using a 5\arcsec\ aperture for the source and a 25\arcsec\ aperture for the background. We subtracted the host fluxes obtained using observations from 2023 March 9, which is during the quiescent phase between the two flares, from all other epochs. 

The location of \dbl\ was observed by the Transiting Exoplanet Survey Satellite \citep[TESS;][]{Ricker2015} in sectors 22, 48, 49, 75 and 76. Sector 22 in Cycle 2 covers the period from MJD 58897 to 58926, i.e., two years before the first flare. Sectors 48 and 49 in Cycle 4 cover MJD 59607 to 59664, which coincides with the first flare. Sectors 75 and 76 in Cycle 6 cover MJD 60339 to 60395, which coincides with the second flare. The position of \dbl\ in Sector 75 is too close to the edge of the imaging array to obtain valid flux calibration, and therefore data from this sector were not used. We processed the TESS data using image subtraction and forced PSF photometry \citep{Fausnaugh2021}. We used a customized version of the ISIS pipeline \citep{AL1998} to perform image subtraction for the science images within each sector with reference images built from the 20 lowest background images in the first seven days of the sector (as cross-sector subtractions are not possible). We used the quiescent phase at the start of Sector 48 (before the first flare) to determine the zero level of the flux from the mean of the photometry and the typical flux errors from the scatter of the photometry during this epoch. We treated any subsequent flux level that is below three times this scatter as a non detection. Since the first seven days of Sector 49 included TDE light, subtracting their photometry from the photometry of the rest of the sector causes an underestimation of the TDE luminosity. Unfortunately, we found the background levels at the end and start of each sector to be too high to extract photometry from those epochs. This leads both to gaps in the TESS lightcurve, and prohibits scaling any over-subtracted sector to the sector before it. We therefore were unable to fully calibrate the TESS photometry, and we do not use it for any modeling purposes. We present it here only for completeness. 

We queried the NASA/IPAC Infrared Science Archive\footnote{\url{https://irsa.ipac.caltech.edu/frontpage/}} for mid-infrared (MIR) detections within 5\arcsec\ of the ZTF-determined position of \dbl. 
This position was visited approximately twice a year by NEOWISE-R \citep{Mainzer2011, Mainzer2014} since MJD 56803. We re-binned the measurements available for each visit (using a weighted mean) into one representative measurement per year per band. We estimated the host-galaxy MIR flux and its uncertainty as the average and variance (respectively) of all pre-TDE epochs and then we subtracted it from all observations. We found no significant MIR activity out to $\sim$2 years after the first peak at the $3\sigma$ level. 

\begin{figure*}
\centering
\includegraphics[width=\textwidth]{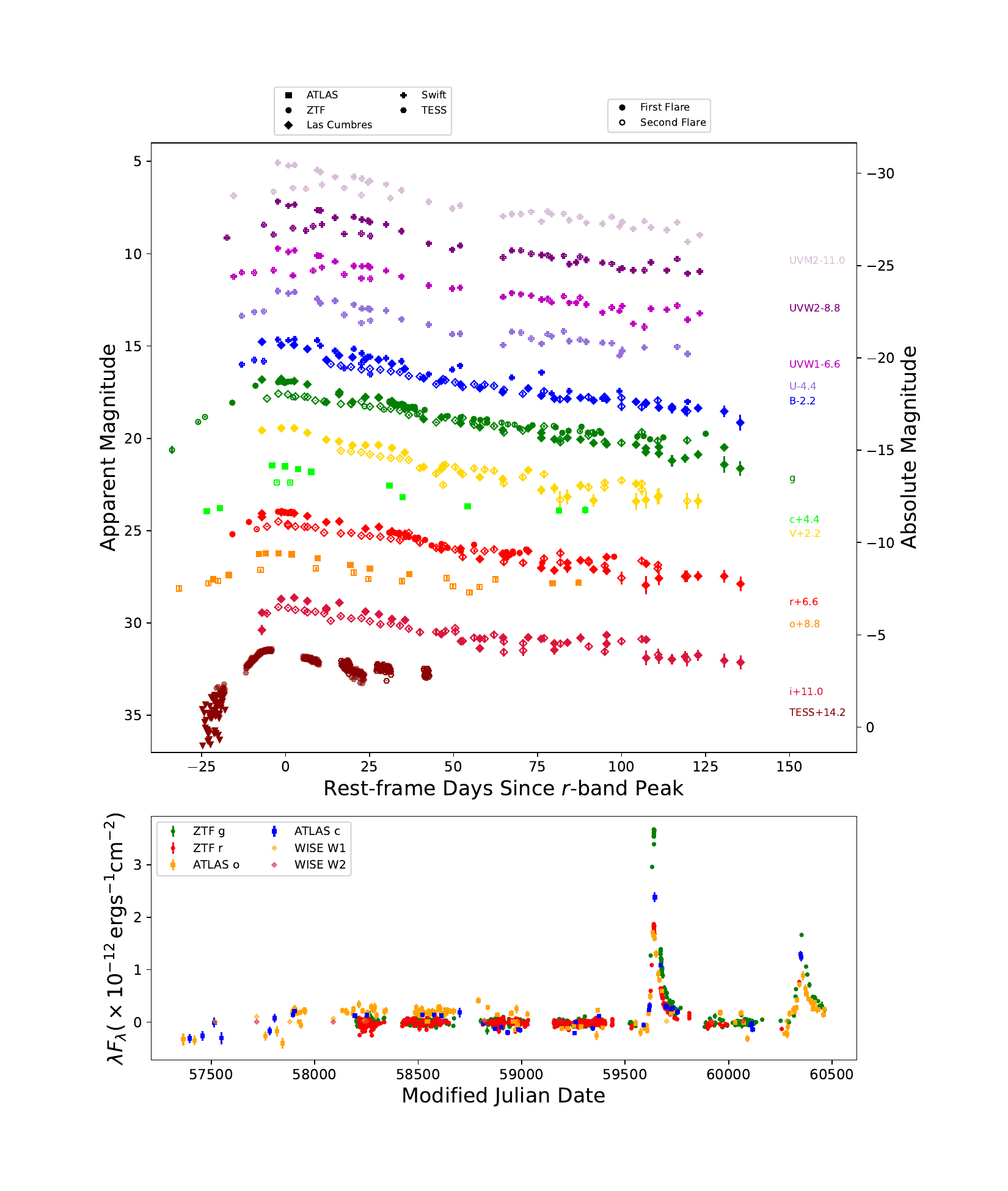}
\caption{Optical and ultraviolet host-subtracted light curves of {\dbl} normalized to each peak time and shifted in magnitude for clarity (both flares are shifted by the same amount per band; top). Both flares are very similar to each other in the optical bands and show some differences in the ultraviolet. No earlier flares were detected in the $\sim$1500 days preceding the first flare (bottom). Error bars denote $1\sigma$ uncertainties and triangles denote $3\sigma$ non-detection upper limits. 
\label{fig:photometry}
}
\end{figure*}

All of our photometry is presented in Figure \ref{fig:photometry} and Table \ref{tab:phot}. The Las Cumbres $B$ and $V$-band photometry are presented in the Vega system, and all other photometry in the AB system.

\begin{deluxetable}{lllll}
    \label{tab:phot}
    \centering
    \caption{{\dbl} photometry.}
    \tablehead{
    \colhead{MJD} & \colhead{Filter} & \colhead{Magnitude} &\colhead{Error} &  \colhead{Source} 
    }
    \startdata
59532.52 & $g$ & $>$20.70 & ... &ZTF\\
59550.47 & $g$ &  $>$20.91 & ... & ZTF \\
59623.33 & $g$  &  18.13 & 0.03 & ZTF  \\      
59630.42 & $g$  & 17.21 & 0.01 & ZTF    \\    
59637.37 & $g$  &  17.02 & 0.01 & ZTF   \\     
59639.32 & $g$  &  17.02 & 0.01 & ZTF   \\     
59671.43 & $g$  &  18.06 & 0.02 & ZTF   \\     
59674.38 & $g$  & 18.23 & 0.03 & ZTF    \\    
........ & ...&....&...&... \\
59729.90 & $r$ &   19.93 & 0.13 & Las Cumbres \\
59733.90 & $r$ &   20.51 & 0.19 & Las Cumbres \\
59737.90 & $r$ &   20.01 & 0.18 & Las Cumbres \\
59750.00 & $r$ &   20.57 & 0.18 & Las Cumbres \\
59632.40 & $i$ &   17.26 & 0.16 & Las Cumbres \\
59642.30 & $i$ &   17.86 & 0.14 & Las Cumbres\\
   \enddata
    \tablecomments{Upper limits denote 3$\sigma$ non-detections. This table is published in its entirety in the machine-readable format. A portion is shown here for guidance regarding its form and content.}
\end{deluxetable}

\subsection{Optical Spectroscopy}

\begin{figure*}
\centering
\includegraphics[width=0.7\textwidth]{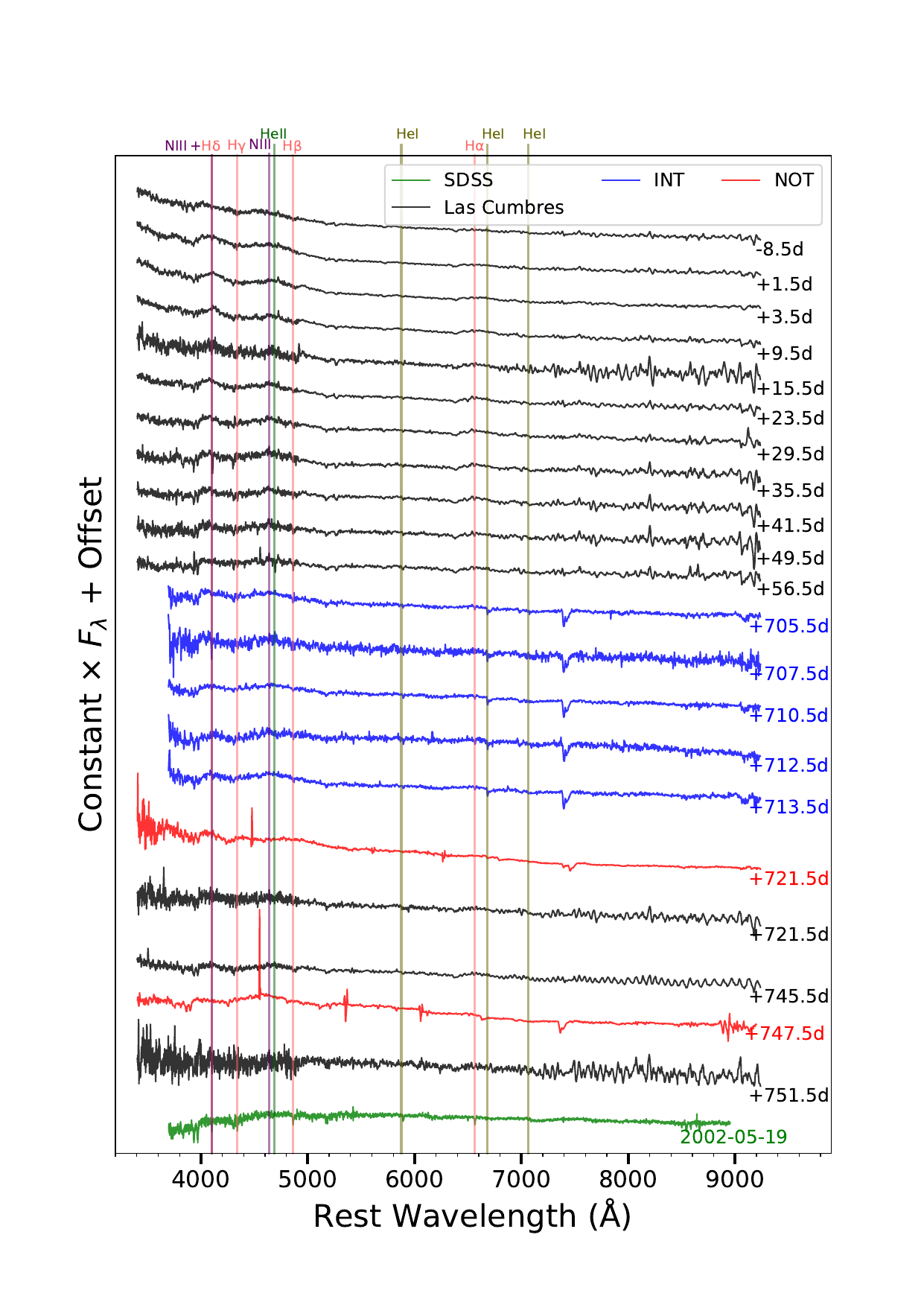}
\caption{Our full spectral series of \dbl\ together with the host-galaxy spectrom from SDSS. Lines that characterize various sub-types of optical-ultraviolet TDEs are marked. Phases are noted in days relative to peak luminosity of the first flare. 
\label{fig:allspec}}
\end{figure*}

We obtained optical spectroscopy of \dbl\ using our program on the FLOYDS spectrograph \citep{Sand2011} mounted on the robotic 2\,m Faulkes Telescope North at Haleakal\=a, Hawaii, which is part of the Las Cumbres Observatory network. FLOYDS covers the 3500--10000\,\AA\ range in a single exposure by capturing two spectral orders simultaneously, with a spectral resolution of R$\sim$500 using a slit width of 2\arcsec. The spectra were reduced with a custom data reduction pipeline\footnote{\url{https://github.com/cylammarco/FLOYDS\_pipeline}} built with the \textsc{iraf}-free \textsc{python}-based \textsc{aspired} toolkit \citep{lam_marco_c_2022_6903357,Lam2023}. Standard data reduction procedures were applied to trace and then optimally extract the spectral information \citep{1986PASP...98..609H}. Wavelength calibration was performed using the built-in calibrator powered by \textsc{rascal} \citep{2020ASPC..527..627V, veitch_michaelis_joshua_2021_4124170}. Standard stars from the same night were used for flux calibration when available; otherwise, standard stars observed closest in time to the science observations were used. Finally, atmospheric extinction and telluric absorption were removed. We obtained five additional optical spectra with the Intermediate Dispersion Spectrograph (IDS) mounted on the 2.54\,m Isaac Newton Telescope (INT) on La Palma, Spain, under the Gravitational-wave Optical Transient Observer \citep{Steeghs2022} Fast Analysis and Spectroscopy of Transients (GOTO-FAST) program. The spectra were obtained using the RED+2 detector with R150V grism with a resolution of R$\sim$600 at 4500\,\AA. The data were reduced using a custom recipe for the PypeIt spectral reduction pipeline \citep{Prochaska2020}. We obtained two additional optical spectra with the Alhambra Faint Object Spectrograph and Camera (ALFOSC) mounted on the Nordic Optical Telescope at La Palma, Spain using the Gr4 grism Gr4 and 1\arcsec\ slit (resulting in a spectral resolution of R$\sim$300). These spectra were reduced using the PyNOT-spex pipeline. We present all spectra in Figure \ref{fig:allspec}. 

\subsection{X-ray Observations}

X-ray observations with the X-Ray Telescope (XRT)  \citep{Burrows2005} on \swift\ were obtained simultaneously with the UVOT observations. Using \texttt{XIMAGE} to process the \swift/XRT images, we found no significant X-ray detection at the position of \dbl. We then used \texttt{XIMAGE} to calculate the corresponding flux upper limits for each of the XRT epochs, using a circular aperture with a radius of $\sim$47\arcsec\ centered on the (optical) position of \dbl\ (we verified that no neighboring X-ray sources are detected within that aperture). Using $\nh=1.94\times10^{20}\,\rm cm^{-2}$, we derive $3\sigma$ upper limits for the absorbed flux across all epochs in the range of $1.1\times10^{-13}\,\ergcms$ to $1.2\times10^{-12}\,\ergcms$, which correspond to $L(0.2-10\kev)<2.0\times10^{41}\,\ergs$. Combining all epochs of the \swift\ XRT data ($\sim$31\,ks exposure), we still do not detect any significant X-ray emission and set a $3\sigma$ upper limit of $1.4\times10^{-14}\,\ergcms$ which corresponds to $L(0.2-10\kev)<2.6\times10^{40}\,\ergs$. 

\subsection{Radio Observations}

\begin{figure}
\centering
% \hspace{-0.8cm}
\includegraphics[width=\linewidth]{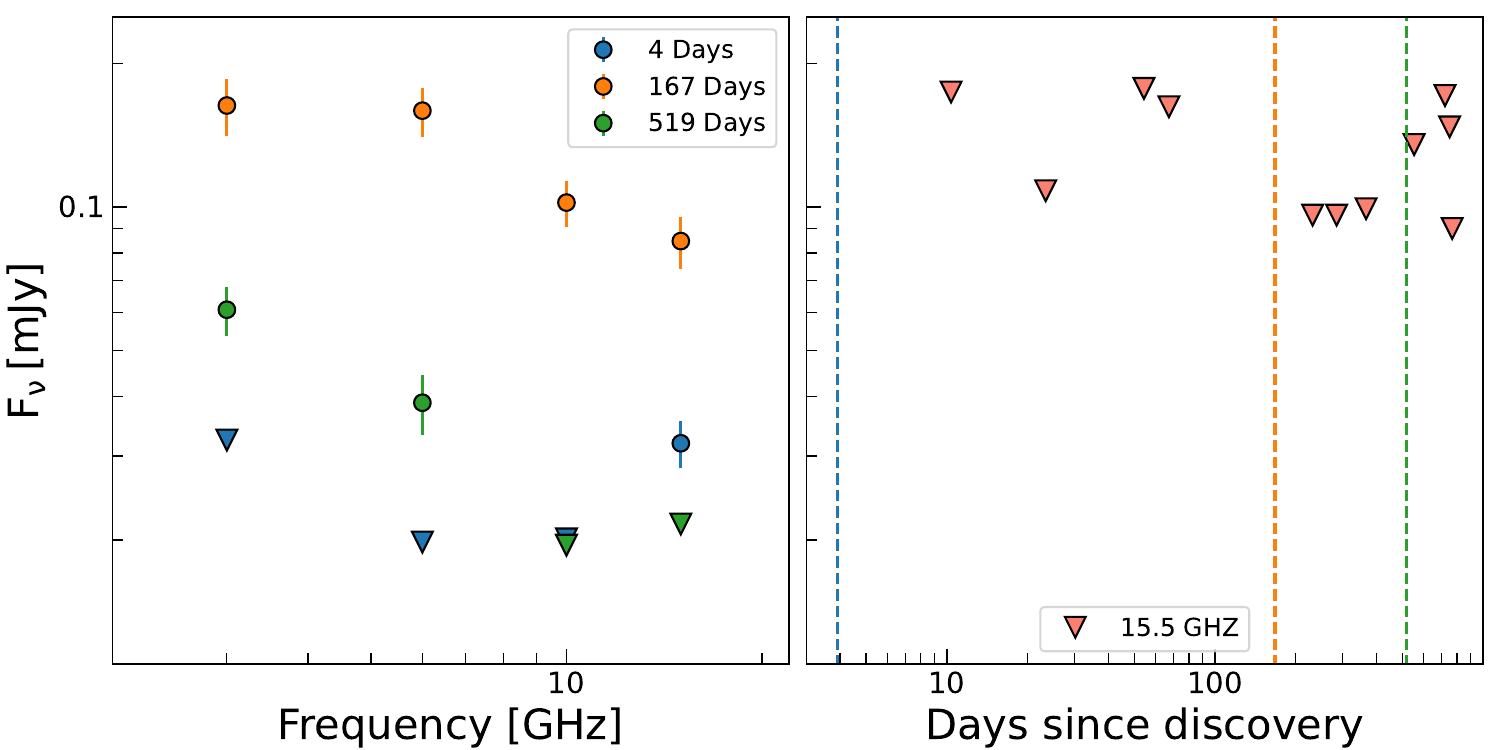}
\caption{Radio spectra at three different epochs (left; denoted in days since optical discovery) and $15.5$ GHz light curve (right) of \dbl, with dashed vertical lines marking the spectral epochs. Error bars denote 1$\sigma$ uncertainties and triangles mark 3$\sigma$ non-detection upper limits.
 \label{EDfig:radio}}
\end{figure}

We observed the position of \dbl\ using the Karl G. Jansky Very Large Array (VLA) under programs 21A-184 and 22A-163 (PI Horesh). Our VLA observations were carried out on MJDs 59636, 59800, and 60152, using the S- C- X- and Ku-bands. The VLA was in BnA$\rightarrow$A configuration during the first observation, in D configuration during the second observation, and in A configuration during the third observation. Automatic flagging and calibration of the data were conducted with the VLA calibration pipeline. We used 3C286 as a primary flux calibrator and J1219+4829 as a phase calibrator. Images of the field of \dbl\ were produced with the CASA tCLEAN task in an interactive mode. When a source was detected at the phase center we fitted it with the CASA task IMFIT, and the image RMS was calculated using the CASA task IMSTAT. We estimate the error of the peak flux density to be a quadratic sum of the error produced by the CASA task IMFIT, and a $10$\,\% calibration error. We detected radio emission consistent with the position of \dbl\ which we present in Figure \ref{EDfig:radio} and Table \ref{tab:radio}. 

We also carried out multiple observations in the $15.5$ GHz band with the Arcminute Micro-Kelvin Imager - Large Array (AMI-LA) \citep{zwart_2008,hickish_2018} from shortly after optical discovery until more than $700$\,days later. Initial flagging and reduction were conducted using $\tt{reduce \_ dc}$, a customized AMI-LA data-reduction software package \citep{perrott_2013}. We use the same primary flux phase calibrators as above. Images of the field of \dbl\ were produced with the CASA CLEAN task in an interactive mode, and the image RMS was calculated using the CASA task IMSTAT. These observations were less sensitive than our VLA observations and resulted in null-detections. Our $3\sigma$ upper limits for the emission from \dbl\ are provided in Table \ref{tab:radio}. 

\begin{deluxetable}{lllll}
    \label{tab:radio}
    \centering
    \caption{{\dbl} radio observations.}
    \tablehead{
    \colhead{MJD} & \colhead{Frequency} & \colhead{$F_{\nu}$} &\colhead{Error} &  \colhead{Source} \\
    \colhead{} & \colhead{(GHz)} & \colhead{(mJy)} &\colhead{(mJy)} &  \colhead{}
    }
    \startdata
$59627$ & $3$ & $\leq 0.033$ & $0.011$ & VLA:BnA$\rightarrow$A \\ 
$59627$ & $6$ & $\leq 0.02$ & $0.006$ & VLA:BnA$\rightarrow$A \\ 
$59627$ & $10$ & $\leq 0.02$ & $0.006$ & VLA:BnA$\rightarrow$A \\ 
$59627$ & $15$ & $0.032 \pm 0.004$ & $0.006$ & VLA:BnA$\rightarrow$A \\ 
$59633$ & $15.5$ & $\leq 0.17$ & $0.058$ & AMI-LA \\
$59646$ & $15.5$ & $\leq 0.11$ & $0.036$ & AMI-LA \\ 
$69677$ & $15.5$ & $\leq 0.18$ & $0.059$ & AMI-LA \\ 
$59690$ & $15.5$ & $\leq 0.16$ & $0.054$ & AMI-LA \\ 
$59790$ & $3$ & $0.16 \pm 0.02$ & $0.020$ & VLA:D \\ 
$59790$ & $6$ & $0.16 \pm 0.02$ & $0.015$ & VLA:D \\ 
$59790$ & $10$ & $0.10 \pm 0.01$ & $0.009$ & VLA:D \\ 
$59790$ & $15$ & $0.085 \pm 0.010$ & $0.015$ & VLA:D \\ 
$59854$ & $15.5$ & $\leq 0.10$ & $0.032$ & AMI-LA \\ 
$59907$ & $15.5$ & $\leq 0.10$ & $0.032$ & AMI-LA \\ 
$59989$ & $15.5$ & $\leq 0.10$ & $0.033$ & AMI-LA \\ 
$60142$ & $3$ & $0.061 \pm 0.007$ & $0.009$ & VLA:A \\ 
$60142$ & $6$ & $0.039 \pm 0.006$ & $0.006$ & VLA:A \\ 
$60142$ & $10$ & $\leq 0.02$ & $0.006$ & VLA:A \\ 
$60142$ & $15$ & $\leq 0.021$ & $0.007$ & VLA:A \\ 
$60176$ & $15.5$ & $\leq 0.14$ & $0.045$ & AMI-LA \\ 
$60344$ & $15.5$ & $\leq 0.17$ & $0.057$ & AMI-LA \\ 
$60373$ & $15.5$ & $\leq 0.15$ & $0.049$ & AMI-LA \\ 
$60390$ & $15.5$ & $\leq 0.09$ & $0.030$ & AMI-LA \\ 
   \enddata
    \tablecomments{Upper limits denote 3$\sigma$ non-detections.}
\end{deluxetable}

\section{Analysis}

\subsection{Optical and Ultraviolet Photometric Analysis}

\subsubsection{Peak Time, Blackbody, and Decline Rate}

We fit a second-degree polynomial around the $r$ band peak of each flare to determine a peak luminosity time of MJD 59639.67$\pm$1.16 for the first flare and MJD 60350.92$\pm$1.09 for the second flare. This gives a time difference of 711.25$\pm$1.59 days between peaks.

\begin{figure}
\centering
\includegraphics[width=\linewidth]{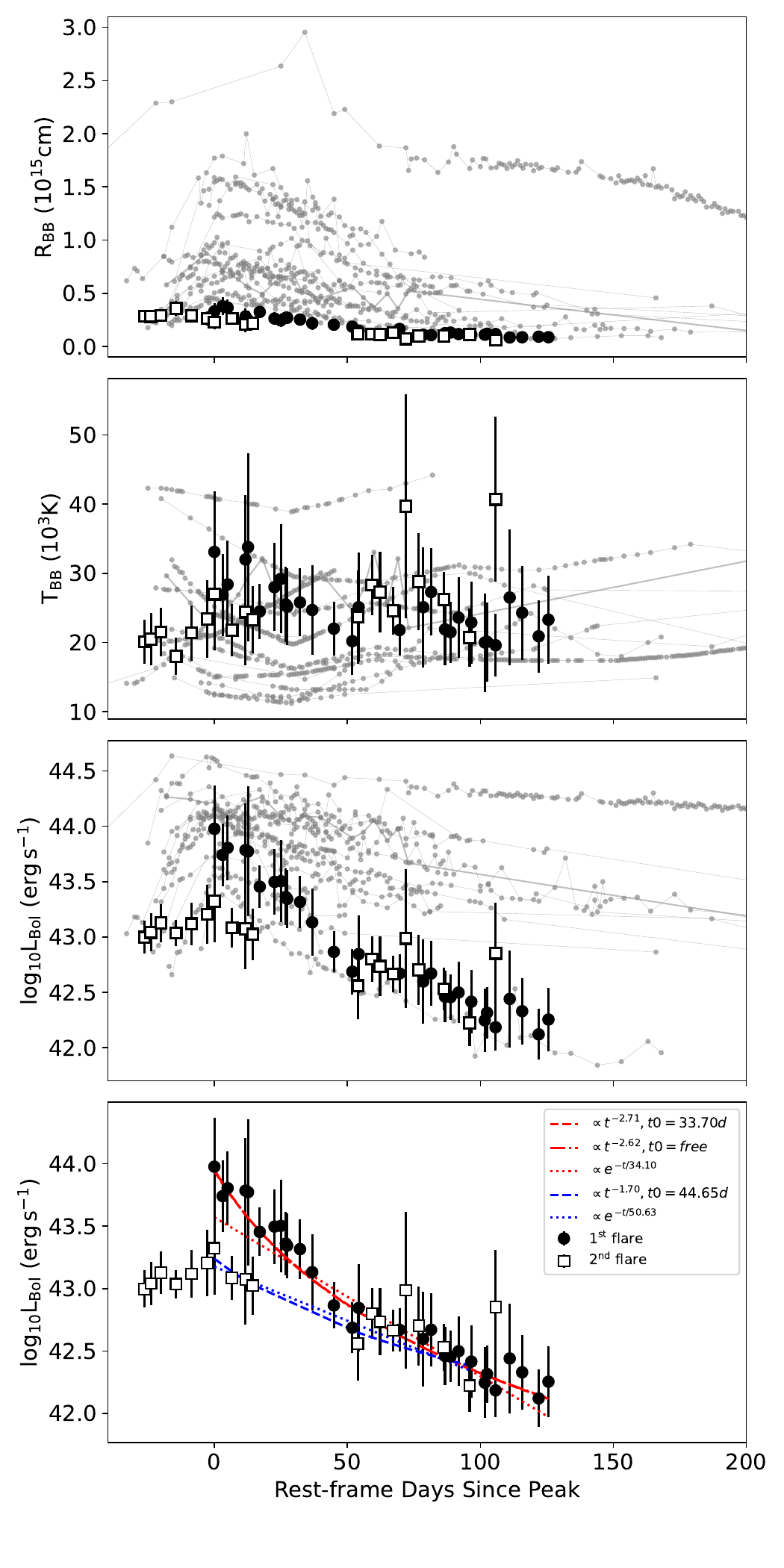}
\caption{\dbl\ best-fit blackbody radius, temperature, and resulting bolometric luminosity (top to bottom panels) (black filled and open markers, for the first and second flares respectively). The temperature and luminosity are within the range of a comparison sample of TDEs \citep[grey;]{VV2020} while the radius is on the lower end of the sample. Fits to the post-peak decline of the bolometric luminosity are also shown (bottom panel). Error bars denote 1$\sigma$ uncertainties and are shown only for \dbl\ for clarity.
\label{fig:bb}}
\end{figure}

We fit a blackbody to the ZTF ($g$ and $r$), ATLAS ($c$ and $o$), Las Cumbres ($g$, $r$, $i$, $B$ and $V$) and \swift\ ultraviolet photometry for each epoch in which we have \swift\ data (linearly interpolating neighbouring optical epochs), using  \textsc{superbol} \citep{Nicholl2018}. We exclude epochs without \swift\ data given the systematic uncertainties introduced when fitting hot blackbodies without ultraviolet observations \citep{Arcavi2022}\footnote{Unfortunately, before peak there are not enough data, even in the optical alone, to constrain blackbody parameters.}. We then calculate the bolometric luminosity by integrating each observed spectral energy distribution, with the missing flux outside the observed bands taken under the assumption of a blackbody with the best-fit parameters. Our results are presented in Figure \ref{fig:bb} and Table \ref{tab:blackbody_fit}. The best-fit temperature and resulting bolometric luminosities of both flares, as well as their evolution in time, are typical of TDEs, while the blackbody radii are on the low end of the known sample. The second flare peak bolometric luminosity is approximately 4 times fainter than that of the first flare. The bolometric peak luminosity of the first (second) flare is $\gtrsim$500 ($\gtrsim$10) times brighter than the per-epoch X-ray non-detection limits, and $\gtrsim$3700 ($\gtrsim$80) times brighter than the stacked X-ray non-detection limit. These values are consistent with blackbody to X-ray luminosity ratios observed in other TDEs \citep{Guolo2024}.

\begin{deluxetable*}{lllllll}
    \label{tab:blackbody_fit}
    \centering
    \caption{{\dbl} best-fit blackbody parameters.}
    \tablehead{
    \colhead{Phase} & \colhead{Temperature} & \colhead{Error} & \colhead{Radius} &  \colhead{Error} & \colhead{Luminosity} & \colhead{Error} \\
    \colhead{(days)} & \colhead{($10^4$K)} & \colhead{($10^4$ K)} & \colhead{({$10^{14}$}cm)} &  \colhead{($10^{14}$cm)} & \colhead{({$10^{43}$}erg\,s$^{-1}$)} & \colhead{({$10^{43}$}erg\,s$^{-1}$)}
    }
    \startdata
0.0&3.31&0.88&3.29&0.80&9.48&8.56\\
3.112&2.69&0.60&3.77&0.85&5.51&3.61\\
4.979&2.84&0.63&3.64&0.80&6.38&4.35\\
11.688&3.20&0.94&2.81&0.77&6.10&5.91\\
12.651&3.38&1.36&2.47&0.89&5.93&8.00\\
17.056&2.45&0.40&3.24&0.57&2.84&1.28\\
22.588&2.80&0.64&2.62&0.59&3.14&2.16\\
24.981&2.92&0.79&2.44&0.65&3.18&2.71\\
26.848&2.55&0.54&2.69&0.59&2.30&1.36\\
27.363&2.52&0.55&2.69&0.62&2.21&1.33\\
32.196&2.58&0.50&2.52&0.50&2.07&1.13\\
36.785&2.47&0.65&2.17&0.61&1.36&9.49\\
44.905&2.20&0.38&2.04&0.42&0.73&0.31\\
51.867&2.02&0.49&1.85&0.53&0.49&0.23\\
54.249&2.51&0.79&1.48&0.49&0.70&0.57\\
69.623&2.18&0.37&1.64&0.32&0.47&0.19\\
78.51&2.51&0.87&1.10&0.40&0.40&0.35\\
81.476&2.73&0.63&1.06&0.25&0.47&0.32\\
86.698&2.19&0.52&1.24&0.34&0.29&0.15\\
88.779&2.15&0.44&1.30&0.32&0.29&0.14\\
91.803&2.36&0.58&1.16&0.32&0.32&0.20\\
96.587&2.29&0.60&1.13&0.33&0.26&0.17\\
101.663&2.00&0.71&1.10&0.48&0.18&0.12\\
102.509&2.01&0.57&1.19&0.41&0.21&0.11\\
105.718&1.96&0.46&1.16&0.33&0.15&0.08\\
111.017&2.65&0.98&0.84&0.32&0.28&0.28\\
115.714&2.43&0.68&0.86&0.26&0.21&0.15\\
121.908&2.09&0.53&0.91&0.28&0.13&0.07\\
125.583&2.33&0.63&0.87&0.26&0.18&0.12\\
677.762&2.01&0.32&2.83&0.55&0.99&0.34\\
680.212&2.05&0.37&2.82&0.62&1.10&0.43\\
683.917&2.15&0.35&2.90&0.56&1.34&0.53\\
689.634&1.80&0.27&3.58&0.70&1.09&0.29\\
695.449&2.14&0.40&2.89&0.63&1.31&0.58\\
701.41&2.34&0.56&2.63&0.69&1.60&0.98\\
704.045&2.70&0.81&2.27&0.69&2.10&1.80\\
710.657&2.18&0.38&2.63&0.54&1.21&0.50\\
715.772&2.44&0.77&2.08&0.74&1.18&0.99\\
718.466&2.33&0.48&2.16&0.51&1.06&0.57\\
757.974&2.37&0.68&1.18&0.37&0.36&0.25\\
763.292&2.83&0.43&1.15&0.18&0.63&0.30\\
766.307&2.73&0.58&1.13&0.25&0.54&0.34\\
771.237&2.46&0.34&1.29&0.20&0.46&0.18\\
776.031&3.97&1.62&0.72&0.25&0.97&1.40\\
780.883&2.88&0.70&0.98&0.24&0.50&0.37\\
790.286&2.62&0.40&0.97&0.16&0.34&0.15\\
800.01&2.07&0.42&1.11&0.28&0.17&0.08\\
    \enddata
    \tablecomments{Phase is given in days relative to the first bolometric peak.}
\end{deluxetable*}

We fit the post-peak bolometric light curve (i.e. the luminosity $L$ vs. time $t$) of each flare with a power law of the form $L=L_0[(t-t_0)/\tau]^{-\alpha}$ and an exponential decay of the form $L=L_0 e^{-t/\tau}$.
For the first flare, we perform the power-law fit twice, once with $t_0$ as a free parameter and once with $t_0$ fixed to 33.7 days before peak (the fallback time of the most bound debris obtained from fitting the lightcurve with a reprocessing-emission TDE model, see below). $L_0$ and $\tau$ are strongly degenerate, and hence are not constrained individually. Here we focus on the power law $\alpha$ which has the most value in constraining physical scenarios. Our best-fit results are presented in Figure \ref{fig:bb} and Table \ref{tab:decline_fit}. The data are better described by a power law than an exponential, with the first flare decline preferring $\alpha=$2.6--2.7, steeper than the canonical $\alpha=5/3$ power law for the mass fallback rate of a full disruption \citep{Rees1988,Phinney1989}. Such steep decline has been seen in a few optical-ultraviolet TDEs \citep[e.g.][]{Nicholl2020,Panos2023}, and is close to expected values for partial disruptions \citep[e.g.]{CN2019}. The decline rate of the second flare is more difficult to constrain given the available data. For fixed $t_0$ at 44.65 days before peak (estimated in the same way as for the first flare), the fit is consistent with a canonical $\alpha=5/3$ decline, but it is not possible to determine if this is part a broad peak which later settled to a steeper decline. We were not able to obtain any constraining fits using a free $t_0$ for the second flare. 

\begin{deluxetable}{ll}
    \label{tab:decline_fit}
    \centering
    \caption{{\dbl} bolometric decline best-fit parameters.}
    \tablehead{
    \colhead{Model} & \colhead{Parameters}
    }
    \startdata
    \multicolumn{2}{c}{Flare 1} \\
    \hline
Power-law (fixed $t_0=-33.70$) & $\alpha=-2.71\pm0.10$  \\
&$\tau$ unconstrained\\
&$L_0$ unconstrained\\
\hline
Power-law (free $t_0$) & $\alpha=-2.62\pm0.45$\\
 & $\tau$ unconstrained \\
 &$t_0=-34.51\pm11.82$\\
 &$L_0$ unconstrained \\
\hline
Exponential & $\tau=34.10\pm2.44$\\
&$L_0=\left(3.72\pm0.56\right)\times10^{43}$\\
    \hline
    \multicolumn{2}{c}{Flare 2} \\
    \hline
Power-law (fixed $t_0=-44.65$) & $\alpha=-1.70\pm0.25$  \\
&$\tau$ unconstrained\\
&$L_0$ unconstrained\\
\hline
Exponential & $\tau=50.63\pm6.02$\\
&$L_0=\left(1.49\pm0.22\right)\times10^{43}$\\
    \enddata
    \tablecomments{$\tau$ and $t_0$ are given in days and $L_0$ in erg\,s$^{-1}$.}
\end{deluxetable}

\subsubsection{Analytical TDE Models}

\begin{figure}
\centering
\includegraphics[trim={10mm, 15mm, 0, 0},clip,width=\linewidth]{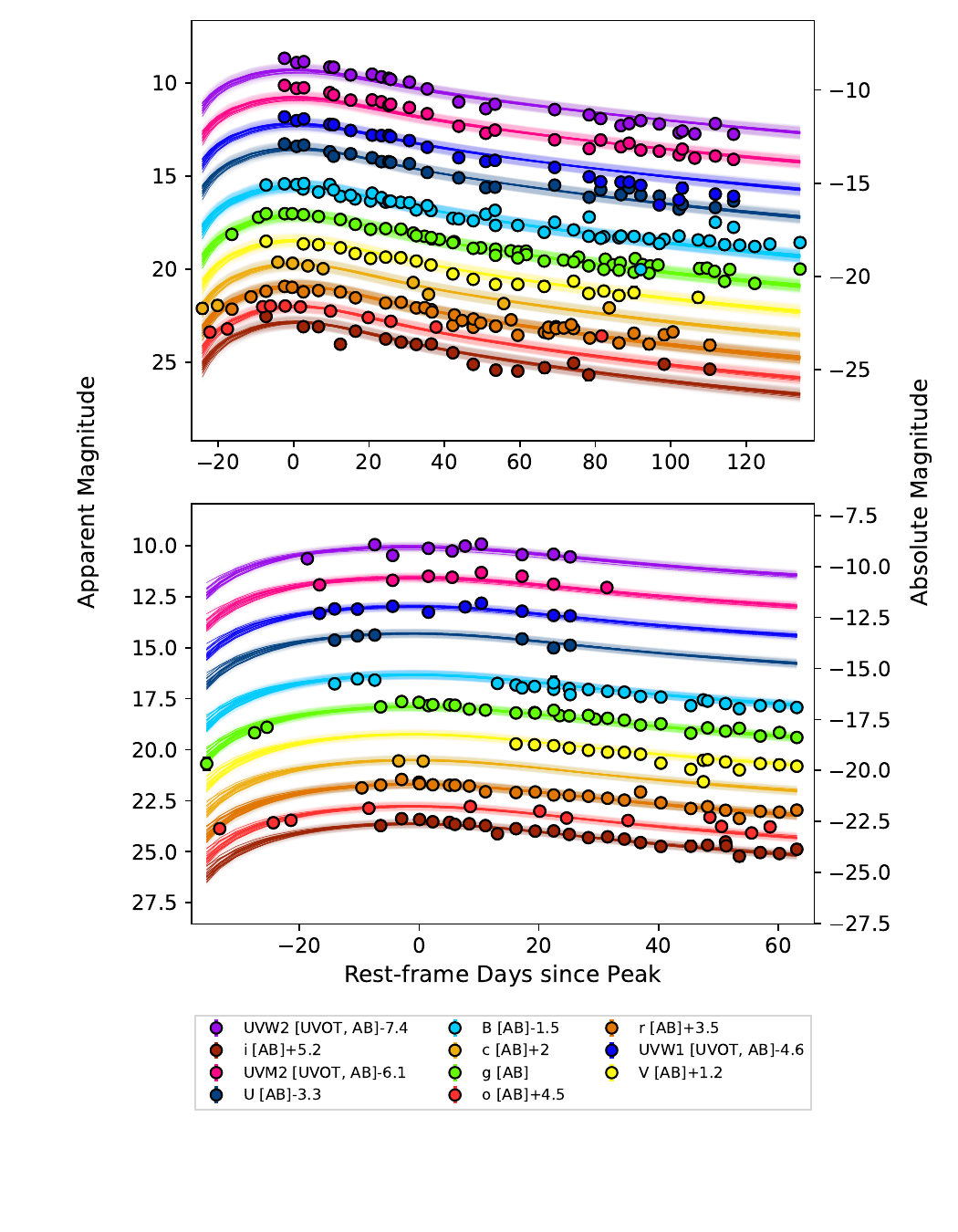}
\caption{Reprocessed accretion emission MOSFiT model fits to the first (top) and second (top) flares of \dbl. 
\label{fig:mosfit}}
\end{figure}

\begin{figure*}
\centering
\includegraphics[width=\textwidth]{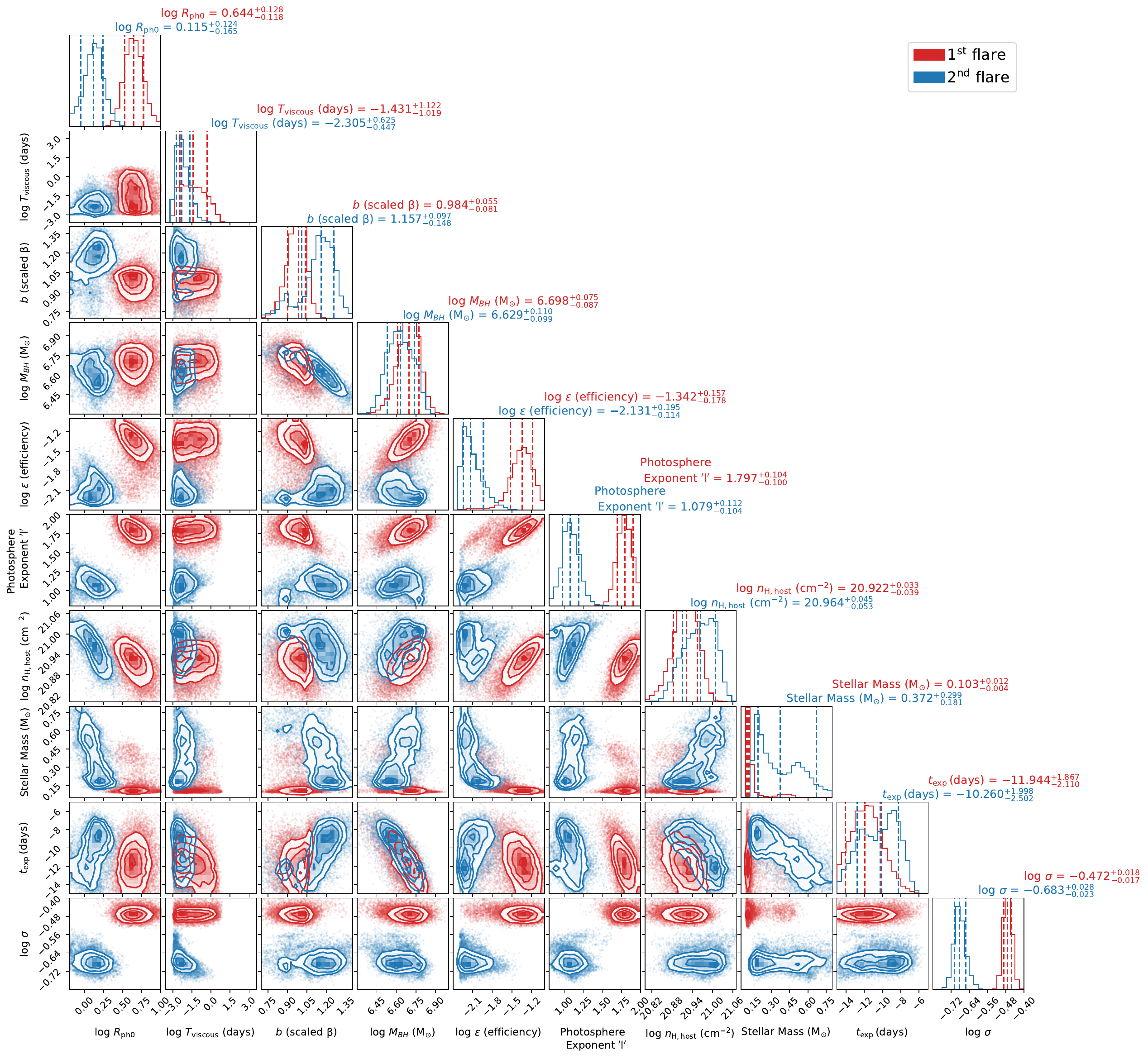}
\caption{Two-dimensional posterior distributions of MOSFiT model fits for both the first (red) and second (blue) flare. While the black hole mass and host-galaxy extinction are consistent for both flares, the $b$ parameter supports both a full and partial disruption for the first flare, while preferring a full disruption for the second flare.
\label{fig:mosfit_corner}}
\end{figure*}

We use the Modular Open Source Fitter for Transients \citep[MOSFiT;][]{Mosfit2018} to fit the multi-band lightcurves of \dbl\ with a TDE emission model \citep{Mockler2019} that assumes a mass fallback rate derived from simulated disruptions \citep{Guillochon2014} of polytropic stars by a black hole of $10^6\,\Msun$. This model then uses scaling relations and interpolations for a range of black hole masses, star masses, and encounter parameters. The free parameters of the model are the mass of the black hole, \mbh, and star, $m_*$; the scaled impact parameter $b$ (which is defined in such way that $b=0$ corresponds to no disruption while $b\geq1$ corresponds to a full disruption); the efficiency $\epsilon$ of converting accreted mass to energy; the unit-less normalization and power law index, $R_{ph,0}$ and $l_{ph}$ respectively, connecting the radius to the instantaneous luminosity; the viscous delay time $T_\nu$ (the timescale for matter to circularize and/or move through the accretion disk) which acts approximately as a low pass filter on the light curve; the time of first fallback, $t_0$; the extinction, proportional to the hydrogen column density \nh\ in the host galaxy; and a white noise parameter, $\sigma$. We use the nested sampling method implemented through \textsc{dynesty} \citep{dynesty}, as appropriate for complex posteriors in high-dimensional parameter spaces. We list the prior distributions for the parameters along with the posteriors in Table \ref{tab:mosfit}. The light curve fits are presented in Figure \ref{fig:mosfit} and the two-dimensional posterior distributions are presented in Figure \ref{fig:mosfit_corner}. We find that the first (second) flare of \dbl\ is best described by the disruption of a $m_*=0.1034^{+0.0121}_{-0.0044}\pm0.6600\,\Msun$ ($m_*=0.37^{+0.30}_{-0.18}\pm0.66\,\Msun$) star by a black hole of mass $\log(\mbh/\,\Msun)=6.698^{+0.076}_{-0.087}\pm0.200$ ($\log(\mbh/\,\Msun)=6.63^{+0.11}_{-0.10}\pm0.20$; statistical and systematic uncertainties \citep{Mockler2019} reported. There is a strong degeneracy between the stellar mass and efficiency parameter in this model \citep{Mockler2021}. The derived parameters are consistent between both flares, and the black hole mass estimate is also consistent with the value found from the galaxy scaling relation described below. The best scaled impact parameter of the first (second) flare is $b=0.984^{+0.054}_{-0.081}\pm0.350$ ($b=1.157^{+0.096}_{-0.148}\pm0.350$). While the scaled impact parameter of the second flare is more consistent with a full disruption, the posterior for the first flare spreads both below and above 1 (Fig. \ref{fig:mosfit_corner}), meaning that it is consistent with both a full and partial disruption. The difference in the photospheric radius normalization factor $R_{ph,0}$ between flares is likely due in large part to the difference in bolometric luminosities, and does not result in large differences to the actual photospheric radius determined by MOSFiT, which varies by a factor of $\lesssim$2 between flares.

\begin{deluxetable*}{lllllll}
    \label{tab:mosfit}
    \centering
    \caption{{\dbl} best-fit MOSFiT parameters and priors used.}
    \tablehead{
    \colhead{Parameter} & \colhead{Prior} & \colhead{Range} &\colhead{Best-Fit Flare 1} &  \colhead{Best-Fit Flare 2} & \colhead{Systematic Error} & \colhead{Unit}
    }
    \startdata
$\mbh$ & Log & [$10^5$, $10^8$] & $6.698^{+0.076}_{-0.087}$&$6.629^{+0.110}_{-0.099}$&$\pm$0.20 & $\Msun$ \\
$m_*$ & Kroupa &[0.01, 100] & $0.1034^{+0.0121}_{-0.0044}$&$0.37^{+0.30}_{-0.18}$&$\pm$0.66 & $\Msun$ \\ 
$b$ & Flat &[0,2] & $0.984^{+0.055}_{-0.081}$& $1.157^{+0.096}_{-0.148}$&$\pm$0.35 &  \\ 
$\epsilon$ & Log & [0.005, 0.4] & $-1.34^{+0.16}_{-0.18}$ &$-2.13^{+0.20}_{-0.11}$&$\pm$0.68&   \\ 
$R_{ph,0}$ & Log & [$10^{-4}$, $10^4$] & $0.64^{+0.13}_{-0.12}$ & $0.12^{+0.12}_{-0.17}$&$\pm$0.40 & \\ 
$l_{ph}$ & Flat & [0, 4] & $1.80^{+0.10}_{-0.10}$&  $1.08^{+0.11}_{-0.10}$&$\pm$0.20 &  \\ 
$T_\nu$ & Log & [$10^{-3}$,$10^3$] & $-1.43^{+1.02}_{-1.12}$& $-2.31^{+0.62}_{-0.45}$&$\pm$0.10 & days\\ 
$t_0$ & Flat & [$-150$, 0] & $-11.94^{+1.87}_{-2.11}$&$-10.26^{+2.00}_{-2.50}$&$\pm$15 & days \\ 
$n_{H,host}$ & Log & [$10^{19}$, $10^{23}$] & $20.922^{+0.033}_{-0.039}$&$20.964^{+0.045}_{-0.053}$ & & $\rm cm^{-2}$ \\ 
$\sigma$ & Log &  [$10^{-4}$, $10^2$] & $-0.472^{+0.018}_{-0.017}$&$-0.683^{+0.027}_{-0.023}$& &
   \enddata
    \tablecomments{Best-fit results are the median of each posterior distribution, and the uncertainties are the 16th and 84th percentiles. These error estimations do not include the systematic uncertainties estimated for MOSFiT \citep{Mockler2019}, shown in a separate column. 'Log' in the Prior column denotes a log-uniform distribution, and values that refer to the log of the parameter.}
\end{deluxetable*}

To estimate the black hole and stellar masses based on the stream collision scenario \citep{Piran2015,Jiang2016}, we use the TDEMass package \citep{Ryu2020b}. This package fits the mass of the black hole and of the disrupted star to the peak bolometric luminosity and the temperature at this time. For the first (second) flare we find a stellar mass of $m_*=0.87_{-0.46}^{+0.39}$ ($m_*=0.40_{-0.30}^{+0.34}$) and a black hole mass of $\log(\mbh/\Msun)=6.00_{-0.36}^{+0.69}$ ($\log(\mh/\Msun)=5.54_{-0.26}^{+1.61}$). Again, both flares are consistent with each other, and the black hole mass is consistent with the one derived from the galaxy scaling relation described below.

\subsubsection{Numerical TDE Model}

As illustrative cases, we repeat hydrodynamical simulations of \cite{Bandopadhyay2024} for two scenarios: The first is the repeated partial disruption of a $3\,\Msun$ evolved star (we do not consider more massive stars given the stellar population age of the host galaxy, see below) by a $10^6\,\Msun$ black hole, for which the binding energy of the star to the black hole was chosen such that the orbital period of the star was $\sim 700$ days and its pericenter radius equal to $R_t$. The second simulation is of the disruption of a $1\,\Msun$ star at zero-age main sequence by a $10^6\,\Msun$ black hole, with a pericenter $\sim R_t/1.4$. 

\begin{figure}
\includegraphics[width=\linewidth]{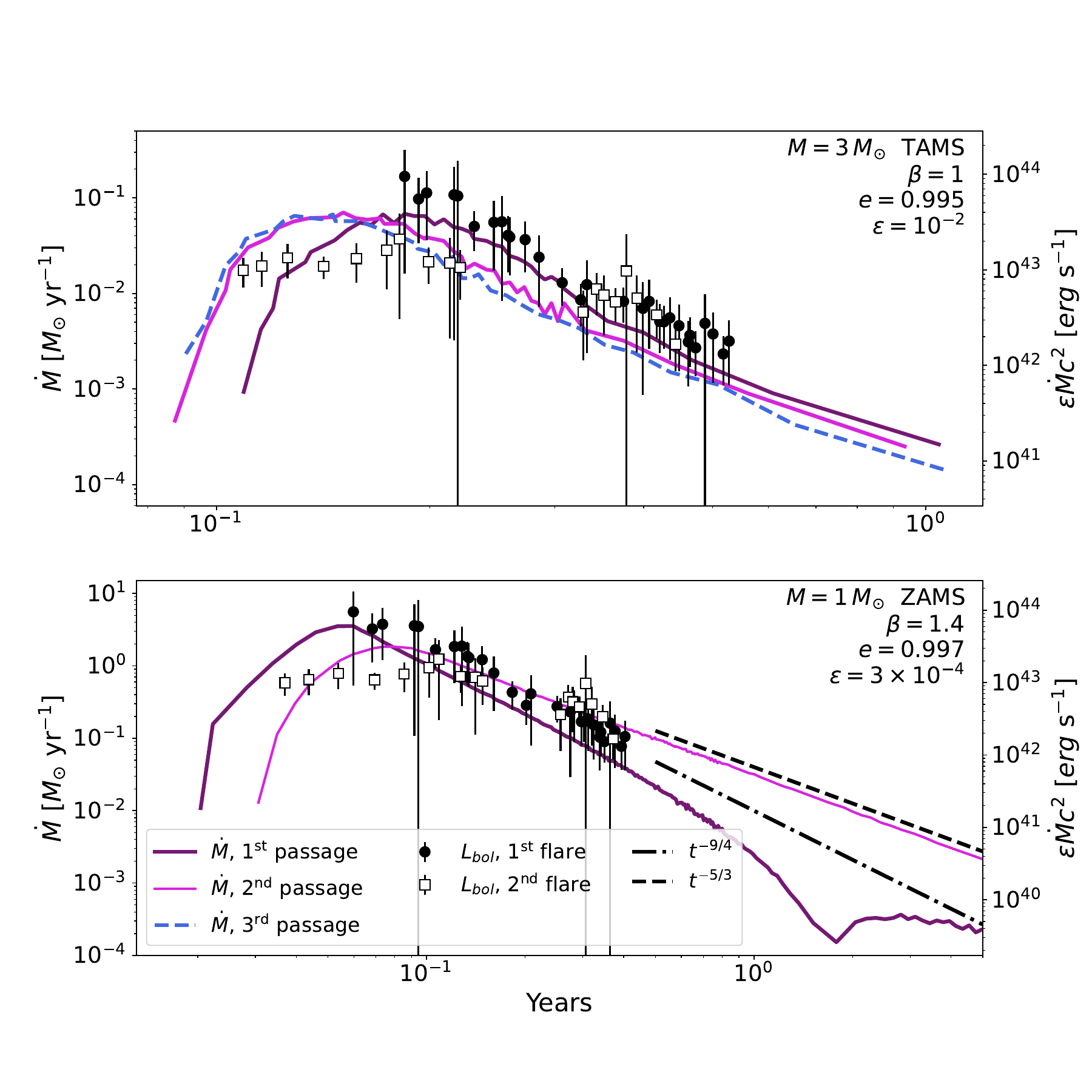}
\caption{Mass fallback rates (left axes) are shown for the first, second and third pericenter passages of a $3\,\Msun$ terminal-age main sequence (TAMS) star on a bound orbit around a $10^6\,\Msun$ black hole, having a pericenter distance $R_{\rm p} \sim R_{\rm t}$ with eccentricity $e=0.995$, consistent with an orbital period of $\sim 700$ days (top), and for the first and second pericenter passages of a $1\,\Msun$ zero-age main sequence (ZAMS) star having a pericenter distance $R_{\rm p} \sim R_{\rm t}/1.4$ with an orbital eccentricity $e=0.997$, consistent with the same orbital period (bottom). In the latter case the star is fully disrupted on the second passage.  The bolometric luminosities of the two flares (right axes) are overlaid with different efficiency parameters $\epsilon$ in each case, to match the mass fallback rate, assuming the luminosity is entirely powered by mass fallback with a constant efficiency. The late-time bump in the $1\,\Msun$ first passage mass fallback rate is due to the less-bound debris returning after one orbital period and is much smaller than the fallback rate due to the bound debris from the second passage. The current data are not able to distinguish between these two types of scenarios. This could be determined by the existence of a third flare. Error bars denote $1\sigma$ uncertainties.
\label{fig:hydrosimulations}
}
\end{figure}

We use the smoothed particle hydrodynamics code {\sc phantom} \citep{Price2018} to simulate the repeated partial disruption of two stars, a 3\,$\Msun$ evolved star and a 1\,$\Msun$ zero-age main sequence star, by a $10^6\,\Msun$ black hole. Using the stellar evolution code {\sc mesa}~ \citep{Paxton2011}, we evolve the stars along the main sequence. The density profiles of the stars are then mapped onto a three dimensional particle distribution in {\sc phantom}, and relaxed for $\sim$5 sound crossing times across the stellar radius \citep{Golightly2019}. We use a resolution of $10^6$ particles to model the stars. The center of mass of the star is placed on a bound orbit around the black hole, with an orbital period of $700$ days, and a pericenter distance $R_{\rm p} \sim R_{\rm t}$ for the 3$\,\Msun$ star, and $R_{\rm p} \sim R_{\rm t}/1.4$ for the $1\,\Msun$ star. The resulting mass fallback rates from the simulations are shown in Figure~\ref{fig:hydrosimulations}. Both scenarios are consistent with the bolometric light curves of \dbl, assuming the bolometric light traces the mass fallback rate. More details can be found in \cite{Bandopadhyay2024}.

In the first simulation, nearly identical mass fallback rates are produced for both flares (and predicted for a third flare; top panel of Figure \ref{fig:hydrosimulations}). In second simulation, the first encounter produces a rapidly declining mass fallback rate, while the second encounter produces a slightly lower mass-fallback peak and subsequently slower decay (bottom panel of Figure \ref{fig:hydrosimulations}). Both scenarios are consistent with the shape of the bolometric light curve of each flare (Figure \ref{fig:hydrosimulations}), though the similarity in the spectral features of the flares (see below) are hard to explain if the two disruptions are significantly different from each other.

These simulations are not fits to the data but are presented as illustrative cases that explore the effect of varying the stellar structure and the pericenter distance on the mass fallback rates from a star on a bound 700-day orbit around a $10^6\,\Msun$ black hole. They are shown with the data only to compare global behavior (such as timescales and decline rates) under the assumption that the fallback of stellar debris onto the black hole is the primary driver of TDE emission. Fitting the observed light curves with the simulated fallback rates would require an exhaustive set of numerical simulations that explores not only the effect of varying stellar structure, but also other parameters such as the SMBH mass, orbital eccentricity and pericenter distance, which is beyond the scope of this work. In addition, it would require a more precise connection between the mass fallback rate and the emission, which is an open problem in TDE physics.

\subsection{Spectroscopic Analysis}

We identify three main broad emission features in the spectra of \dbl\ (Figure \ref{fig:allspec}). The first, around 4100\,\AA, we attribute to $\niii\,\lambda\lambda4097,4103$, possibly from the Bowen Fluorescence mechanism \citep{Bowen1928,Netzer1985}, as seen in other optical-ultraviolet TDEs \citep{Blagorodnova2019,Leloudas2019}. As argued previously \citep{Leloudas2019}, an association with \Hdelta\ is less likely given a lack of (or very weak) H$\gamma$. Next, there is a very broad (and likely blended) feature from $\sim$4300\,\AA\ to $\sim$5000\,\AA\, which encompasses \hbeta, \HeIIop\, and \NIII. Last is the broad feature corresponding to \halpha. In addition, possible weak broad \hei\ emission at 5876\,\AA\ can be seen after host and continuum subtraction (see below). These features classify \dbl\ as a ``Bowen-TDE'' \citep{VV2020}. 
We identify these features both during the first and the second flare, which exhibit nearly identical spectra. 

\begin{figure}
\centering
\includegraphics[width=\linewidth]{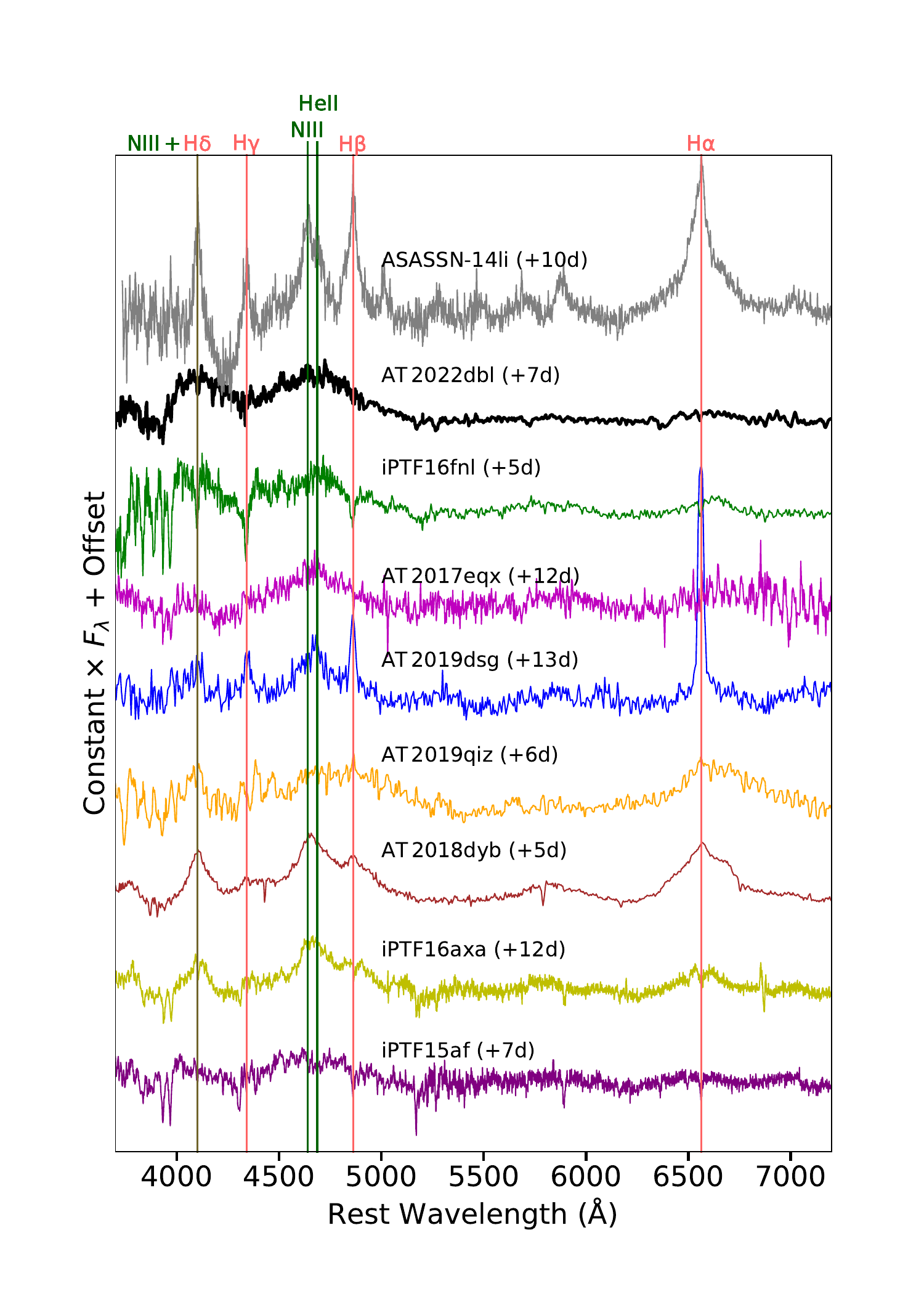}
\caption{Spectroscopic comparison of \dbl\ with a sample of Bowen-TDEs in similar phases with respect to peak luminosity. The spectra have been continuum subtracted to highlight the emission features. \dbl\ is clearly a member of the Bowen TDE class.
\label{EDfig:bowen}}
\end{figure}

In Figure \ref{EDfig:bowen} we compare the spectrum of \dbl\ near peak of its first flare to the near-peak spectra of the Bowen TDEs ASASSN-14li \citep{Holoien2016}, iPTF15af \citep{Blagorodnova2019}, iPTF16axa \citep{Hung2017}, iPTF16fnl \citep{Blagorodovna2017,Onori2019}, AT\,2017eqx \citep{Nicholl2019}, AT\,2018dyb \citep{Leloudas2019}, AT\,2019dsg \citep{Cannizzaro2021} and AT\,2019qiz \citep{Nicholl2020}. While different events show different line profiles, the main features (namely, broad Balmer series, \heii\ and \niii\ emission lines) are common to all events. 

We further analyze the spectra using a standard procedure \citep{Panos2022}, which includes scaling the spectra to the $gri$-band photometry, correcting for Milky Way extinction, subtracting the host-galaxy spectrum from each TDE spectrum, and removing the continuum by fitting line-free regions (3800 -- 3900\,\AA, 4300 -- 4400\,\AA, 5100 -- 5500\,\AA, 6000 -- 6300\,\AA, and 7100 -- 7300\,\AA) with a fourth-degree polynomial. %Normalized regions of interest in the spectra after host and continuum subtraction are presented in Figure \ref{fig:normspectra}. 
As the emission lines in the blue part of the spectrum are likely a blend of various species, we focus our spectroscopic analysis on the \halpha\ emission line. We use the \textsc{lmfit} package to fit the line with a Gaussian profile and calculate the total integrated flux of the Gaussian and its full width at half maximum (FWHM) for each spectrum. The uncertainties for the flux and line width are the 1$\sigma$ uncertainty from the least squares method used for the fitting. 

\begin{figure}
\centering
\includegraphics[width=\linewidth]{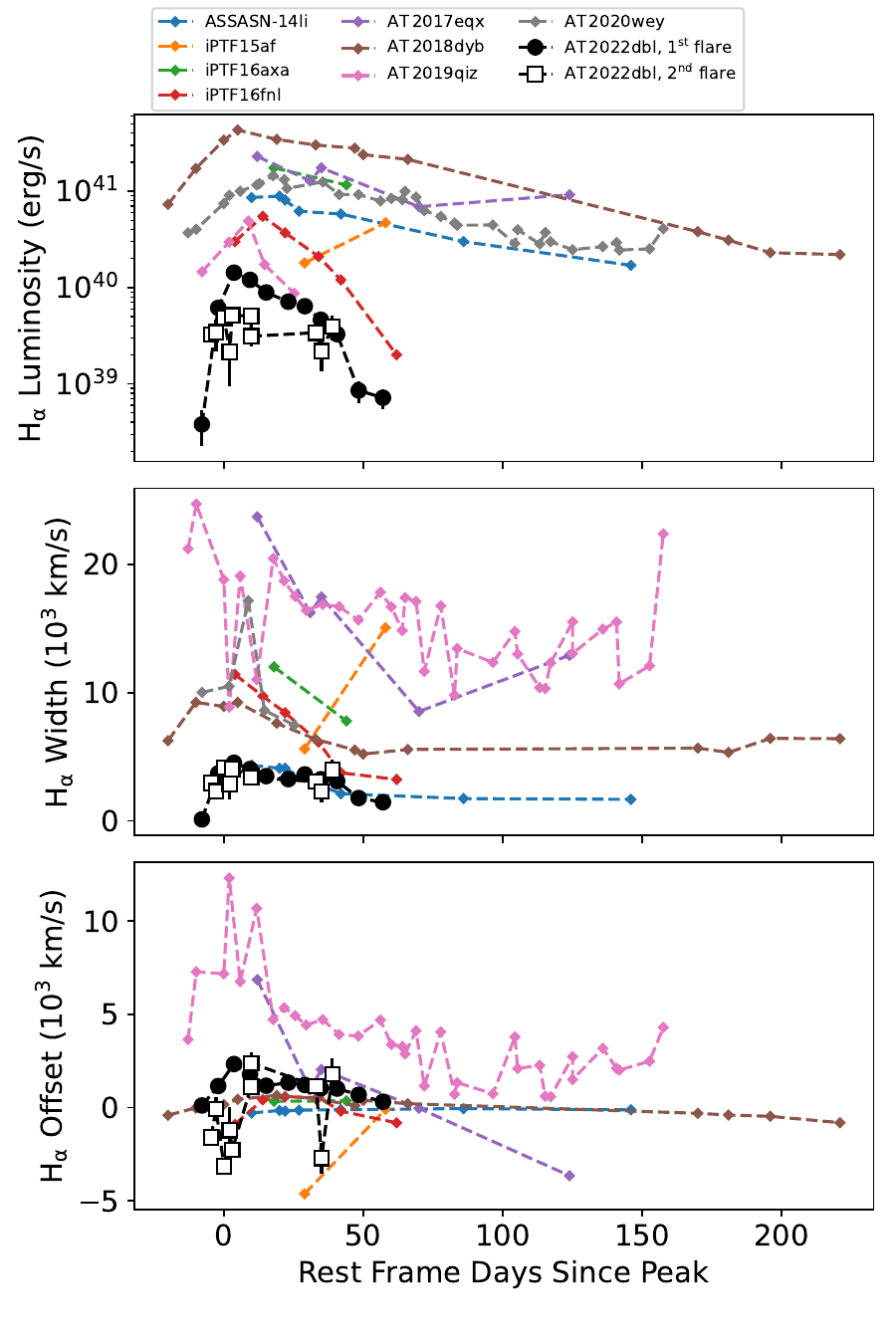}
\caption{\halpha\ luminosity (top), FWHM (middle) and offset (bottom) evolution of \dbl, compared with those of a sample of TDEs. \dbl\ has the lowest \halpha\ luminosity and width of the sample consistently in both flares. Both flares are nearly identical in the \halpha\ width. 
\label{EDfig:halpha}}
\end{figure}

We present the results in Figure \ref{EDfig:halpha} and compare them with a similar analysis performed by \cite{Nicholl2020} and \cite{Panos2022,Panos2023} for the well-sampled Bowen TDE's ASASSN-14li \citep{Holoien2016}, iPTF15af \citep{Blagorodnova2019}, iPTF16axa \citep{Hung2017}, iPTF16fnl \citep{Blagorodovna2017,Onori2019}, AT\,2017eqx \citep{Nicholl2019}, AT\,2018dyb \citep{Leloudas2019}, and AT\,2019qiz \citep{Nicholl2020}, as well as the non-Bowen but rapidly declining TDE AT\,2020wey \citep{Panos2023}. \dbl\ has the weakest measured \halpha\ emission luminosity and lowest FWHM in the sample, perhaps representing a transition between H-rich and H-poor TDEs. The \halpha\ feature evolves similarly in both \dbl\ flares, compared to the spread in the comparison sample. %, indicating that it is very unlikely that both flares are unrelated events.

We find a time lag between the peak of the \halpha\ luminosity (determined using the brightest point) and that of the $g$-band lightcurve of $\sim$5.6$\pm$0.4 days and $\sim$4.8$\pm$2.1 days for the first and second flares, respectively. This timescale corresponds to a light-travel distance of (1.2--1.5)$\times 10^{16}$ cm, which is two orders of magnitude larger than the blackbody radius of \dbl\ (where the continuum emission is formed). Similar results were found for other TDEs \citep{Panos2022}. 

\subsection{Radio Analysis}

The early radio emission observed with the VLA $4$ days after optical discovery of the first flare revealed a faint $0.03$ mJy source in the Ku band and null detections in the S- C- and X-bands (left panel of Figure \ref{EDfig:radio}). The second spectra obtained $167$ days after optical discovery showed an optically thin emission at frequencies higher than $6$ GHz, and a possible transition to the optically thick regime around $3$ GHz, with a peak flux of $\sim 0.16$ mJy. Finally, the last broadband spectrum, obtained $519$ days after optical discovery revealed optically thin emission at low frequencies ($3{-}6$ GHz) and null detections at higher frequencies, suggesting an optically thin spectrum at all GHz frequencies. The $15.5$ GHz light curve obtained with the AMI-LA resulted in multiple null detections from $10$ days to $\sim$2 years after optical discovery, at a $3\sigma$ upper limit of $0.1-0.2$ mJy (right panel of Figure \ref{EDfig:radio}). 

Under an equipartition analysis \citep{barniol_duran_2013}, and using the lowest frequency observed $167$ days after optical discovery as an estimation of the peak flux density, we next estimate the physical parameters of the shock and its environment at that time. We assume that the spectral peak frequency is the synchrotron self-absorption frequency to estimate a radius for the emitting region of $\sim 2 \times 10^{16}$ cm and a minimal total energy of $\sim 3 \times 10^{47}$ erg, respectively. Here we used $\epsilon_{\rm e} = \epsilon_{\rm B} = 0.1$ (where $\epsilon_{\rm e}$ and $\epsilon_{\rm B}$ are the fractions of energy deposited in the relativistic electrons and the magnetic fields, respectively), an area filling factor of $f_{\rm A}=1$, a volume filling factor of $f_{\rm V} = 4/3$, and an electron power-law index of $p=2.5$. Assuming free expansion of the forward shock we find a shock velocity of $\sim 13,000 \rm \, km \, s^{-1}$. We estimate the external density to be $\sim 70 \, \rm cm^{-3}$ by dividing the number of emitting electrons by $4 V$ where $V$ is the volume of the emitting region \citep[a factor of $4$ is needed to account for the shock jump conditions;][]{barniol_duran_2013}. 

Since the transition to the optically thick regime is not observed, the analysis above should be taken with care as the uncertainties on the spectral peak are large. The spectrum obtained $167$ days after discovery is the only one exhibiting the spectral peak, and therefore we do not use other spectra to estimate the physical parameters. However, the spectrum obtained $4$ days after optical discovery shows optically thick emission at $15.5$ GHz and therefore a spectral peak at higher frequencies, and the spectrum at $519$ days shows optically thin emission at $3$ GHz and therefore a spectral peak at lower frequencies. This evolution of the spectral peak to lower frequencies is in agreement with the spectral evolution expected from a shock traveling in a declining density profile \citep{Chevalier_1998}. 

Finally, while the $15.5$ GHz light curve is not sensitive enough for the detection of the first flare we are able to rule out a second, delayed, radio flare \citep[as seen in some optical-ultraviolet in TDEs;][]{horesh_2021a,horesh_2021b,sfaradi_2022,cendes_2022,sfaradi_2024} at luminosities $(3{-}6) \times 10^{37} \, \rm erg \, s^{-1}$ during the first two years from optical discovery.

\subsection{Host Galaxy}

The host galaxy of \dbl\ was classified as a quiescent Balmer-strong galaxy \citep{French2018} based on its SDSS spectrum, having an SDSS $\halpha$ equivalent width in emission of $-0.016\pm0.130$\,\AA\ and an SDSS Lick H$\delta_{A}$ index of $2.20\pm0.58$\,\AA. The MPA-JHU DR8 catalogue lists a velocity dispersion of $\sigma=60.0\pm5.1\kms$ for the galaxy. Using the $\mbh-\sigma$ relation \citep{KH2013}, we find a central black hole mass of $\log(\mbh/\Msun)=6.20\pm0.52$.

\begin{figure}
\centering
\includegraphics[width=\linewidth]{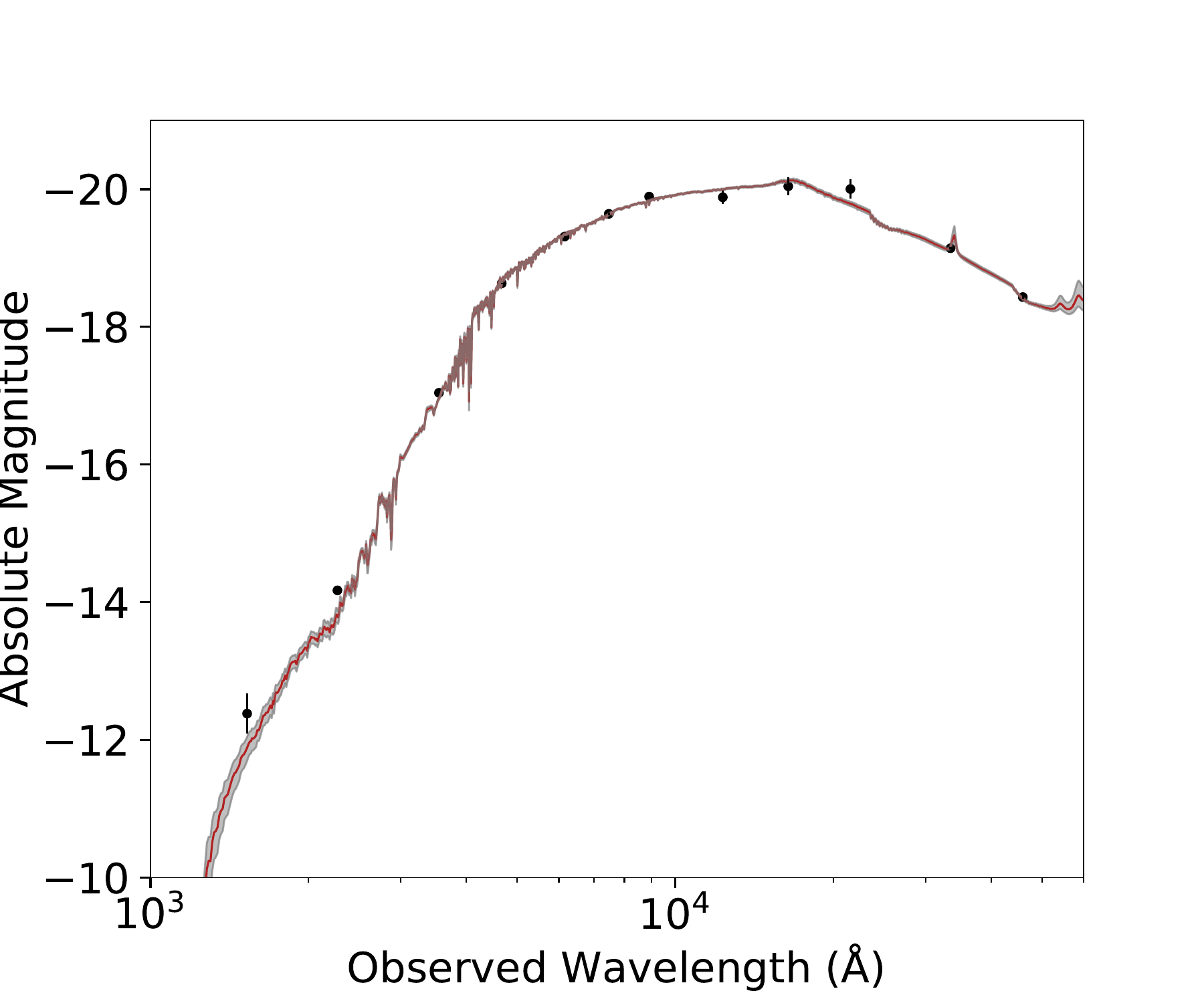}
\caption{Our \textsc{prospector} best-fit median model and 16th and 84th percentile ranges for the \dbl\ host-galaxy spectral energy distribution.
\label{EDfig:host}}
\end{figure}

We retrieved $ugriz$ photometry of the host galaxy from SDSS DR 15 \citep{SDSSDR15}, $JHK$ photometry from 2MASS \citep{twomass2000}, and $W1, W2$ photometry from WISE \citep[through the
AllWISE catalog;][]{Cutri2014} using coadded images taken from 2010 May 25 to 2010 December 3 via the SDSS SkyServer and the NASA/IPAC Extragalactic Database (NED). We also retrieved ultraviolet archival photometry from GALEX \citep{galex2012}. We fit the host-galaxy photometry using the \textsc{prospector} \citep{Leja2017} $\alpha$ model, similar to what was done in previous TDE host-galaxy studies \citep{Nicholl2020,Ramsden2022}. The free parameters in the model are the stellar mass, stellar metallicity, a six-parameter star formation history, and three parameters that control the dust fraction and reprocessing. The observed host-galaxy spectral energy distribution and the best-fit \textsc{prospector} models are shown in Figure \ref{EDfig:host}. We find a stellar mass of $\log(\mstar/\Msun)=10.36^{+0.02}_{-0.03}$, a subsolar metallicity of $\log(Z/\Zsun)=-1.20^{+0.30}_{-0.28}$ and a low specific star-formation rate of $\log(sSFR)=-12.56_{-0.66}^{+0.41}$ within the past 100 Myr (median and 16th and 84th percentiles of the marginalized posterior distributions are given). The stellar mass reported by \textsc{prospector} is the integral of the star formation history and so includes both stars and stellar remnants.

\section{Discussion}

\subsection{\dbl\ Compared to Other TDEs}

\begin{figure}
\centering
\includegraphics[width=\linewidth]{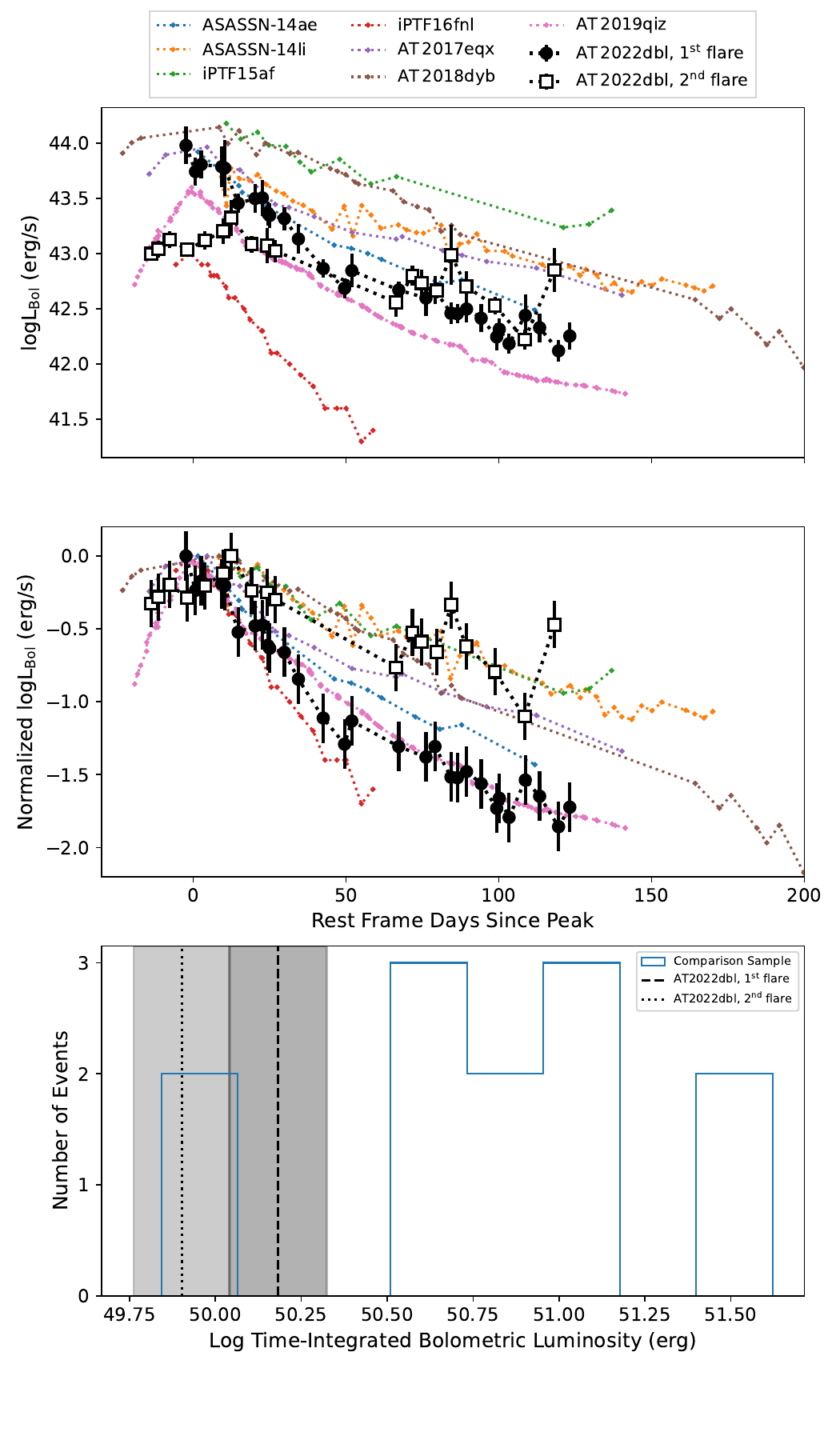}
\caption{Compared to a sample of other Bowen-TDEs, both flares of \dbl\ show typical peak bolometric luminosities (top) and decline rates (middle; same as top panel but normalized to the brightest point). The second flare of \dbl\ is fainter and shallower than the first. Error bars denote $1\sigma$ uncertainties. We also present the time-integrated bolometric luminosity (bottom) of each flare of {\dbl} (dashed lines) compared to those of optical-ultraviolet TDEs from the literature (see Table \ref{tab:intenergy}). Here we use only events with data between peak (or earlier) and 110 days post peak, and we integrate between the first available point out to 110 days. \dbl\ shows relatively low emitted energy, but within the range spanned by the comparison sample. Grey bands denote $1\sigma$ uncertainties around the \dbl\ values. All bolometric light curves are calculated from best-fit blackbody parameters.
\label{fig:lbol}
}
\end{figure}

In the top panels of Figure \ref{fig:lbol} we compare the bolometric light curve of \dbl\ with those of the Bowen TDEs ASASSN-14li \citep{Holoien2016}, iPTF15af \citep{Blagorodnova2019}, iPTF16fnl \citep{Blagorodovna2017,Onori2019}, AT\,2017eqx \citep{Nicholl2019}, AT\,2018dyb \citep{Leloudas2019} and AT\,2019qiz \citep{Nicholl2020}, and with the well-sampled non-Bowen optical-ultraviolet TDE ASASSN-14ae \citep{Holoien2014}. All bolometric light curves are derived from the best-fit blackbody parameters of each event. \dbl\ has typical peak bolometric luminosities among Bowen TDEs, with decline rates consistent with the comparison sample. 

In the bottom panel of Figure \ref{fig:lbol} and in Table \ref{tab:intenergy}, we compare the total radiated energy of each flare of \dbl\ between the earliest available data point (but no later than peak) and 110 days after peak to those of TDEs with similar data coverage: PS1-10jh \citep{Gezari2012}, AT\,2017eqx \citep{Nicholl2019}, AT\,2018dyb \citep{Leloudas2019}, AT\,2018hco \citep{VV2021}, AT\,2018hyz \citep{VV2021}, AT\,2018iih \citep{VV2021}, AT\,2018zr \citep{VV2021}, AT\,2019azh \citep{Faris2024}, AT\,2019dsg \citep{VV2021}, AT\,2019ehz \citep{VV2021}, AT\,2019eve \citep{VV2021}, and  AT\,2019qiz \citep{Nicholl2020} on the same time range. While on the low side of emitted energies, each flare of \dbl\ is still within the range spanned by the comparison sample. 

The top panel of Figure \ref{fig:bb} shows that the blackbody radii of both flares of \dbl\ are at the extreme low end of the comparison sample. It is not known what sets the blackbody radius in optical-ultraviolet TDEs. According to the reprocessing emission picture, this radius could be related to the size of the reprocessing layer of material, however the origin of the reprocessing material itself is not well understood. It is therefore not clear how to interpret the low inferred radii until more such events are discovered.

\begin{deluxetable*}{llll}
    \label{tab:intenergy}
    \centering
    \caption{Integrated bolometric energy of TDEs plotted in the bottom panel of Figure \ref{fig:lbol}.}
    \tablehead{
    \colhead{Name} & \colhead{Time Range} & \colhead{Integrated Energy} &\colhead{Reference} \\
    \colhead{} & \colhead{(days)} & \colhead{(log(erg\,s$^{-1}$))} &\colhead{}
    }
    \startdata
PS1-10jh & [-78.6, 110.0] & $51.6^{+0.1}_{-0.1}$& \cite{Gezari2012}\\
AT2017eqx & [-14.5, 110.0] & $50.5^{+0.1}_{-0.1}$ & \cite{Nicholl2018}\\
AT2018dyb & [-23.1, 110.0] & $50.9^{+0.1}_{-0.1}$ & \cite{Leloudas2019}\\
AT2018hco & [-22.0, 110.0] & $51.1^{+0.1}_{-0.1}$ & \cite{VV2020}\\
AT2018hyz & [0.0, 110.0] & $50.7^{+0.1}_{-0.1}$ & \cite{VV2020}\\
AT2018iih & [-42.0, 110.0] & $51.5^{+0.1}_{-0.1}$ & \cite{VV2011}\\
AT2018zr & [-23.0, 110.0] & $50.6^{+0.1}_{-0.1}$ &  \cite{VV2020}\\
AT2019azh &  [-18.0, 110.0] & $51.0^{+0.1}_{-0.1}$ &  \cite{Faris2024}\\
AT2019dsg & [-20.0, 110.0] & $51.1^{+0.1}_{-0.1}$ &  \cite{VV2020}\\
AT2019ehz & [-16.0, 110.0] & $50.8^{+0.1}_{-0.1}$ & \cite{VV2020}\\
AT2019eve & [-5.0, 110.0] & $49.8^{+0.1}_{-0.1}$ & \cite{VV2020}\\
AT2019qiz & [-19.2, 110.0] & $50.0^{+0.1}_{-0.1}$ &  \cite{Nicholl2020}\\
AT2022dbl (Flare 1) & [0.0, 110.0] & $50.2^{+0.1}_{-0.1}$ &This work\\
AT2022dbl (Flare 2) & [-22.8, 110.0] & $49.9^{+0.1}_{-0.1}$ &This work\\
   \enddata
    \tablecomments{The integration time range is given in rest-frame days relative to peak.}
\end{deluxetable*}

\begin{figure}
\centering
\includegraphics[width=\linewidth]{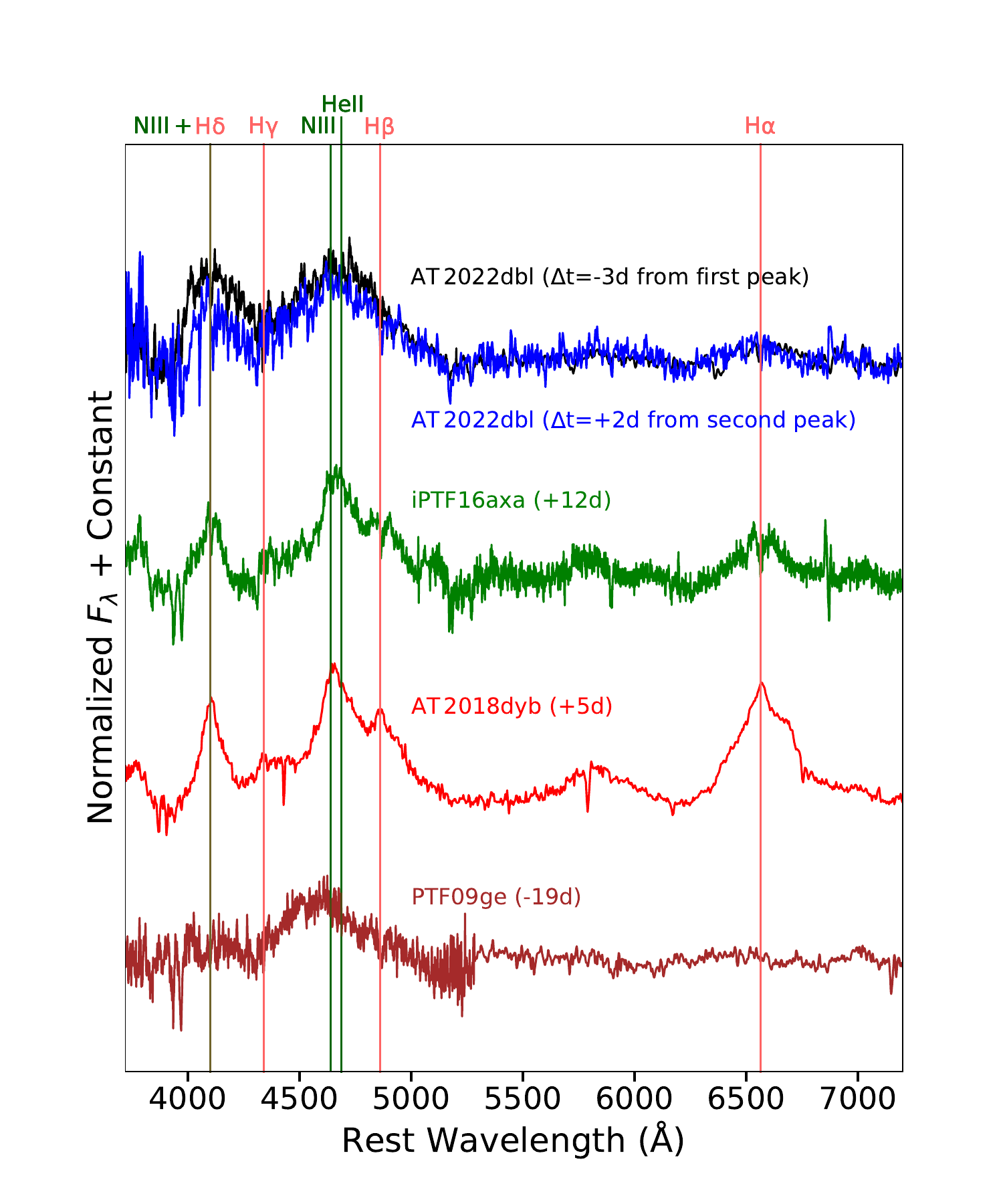}
\caption{Continuum subtracted spectra of \dbl\ near peak luminosity of each flare compared to those of optical-ultraviolet TDEs from the literature. Phases are noted in rest-frame days relative to peak luminosity. The spectral features seen during both flares of \dbl\ are virtually identical to each other compared to the variability seen in the TDE population (especially regarding the \halpha\ strength). 
\label{fig:spectra}
}
\end{figure}

Further to the spectral comparison of Figure \ref{EDfig:bowen}, which places \dbl\ in the He-rich Bowen-TDE class \citep{VV2020}, we compare the spectra near peak luminosity of each flare of \dbl\ to those of the tidal disruption events PTF09ge \citep{Arcavi2014}, which shows a similar \heii\ emission profile and no \halpha\ emission, and with iPTF16axa \citep{Hung2017} and AT\,2018dyb \citep{Leloudas2019}, which show similar \niii\ emission profiles and varying \halpha\ strengths, in Figure \ref{fig:spectra}. This comparison demonstrates how the similarity between the spectra of both flares of \dbl\ is striking compared to the ranges of spectra seen in TDEs. Specifically, both flares of \dbl\ exhibit identically weak \halpha, in between the \halpha\ strengths seen in other TDEs,

\subsection{Possible Origins of the Double Flare in \dbl}

\subsubsection{Gravitational Lensing}

\begin{figure}
\centering
\includegraphics[width=\linewidth]{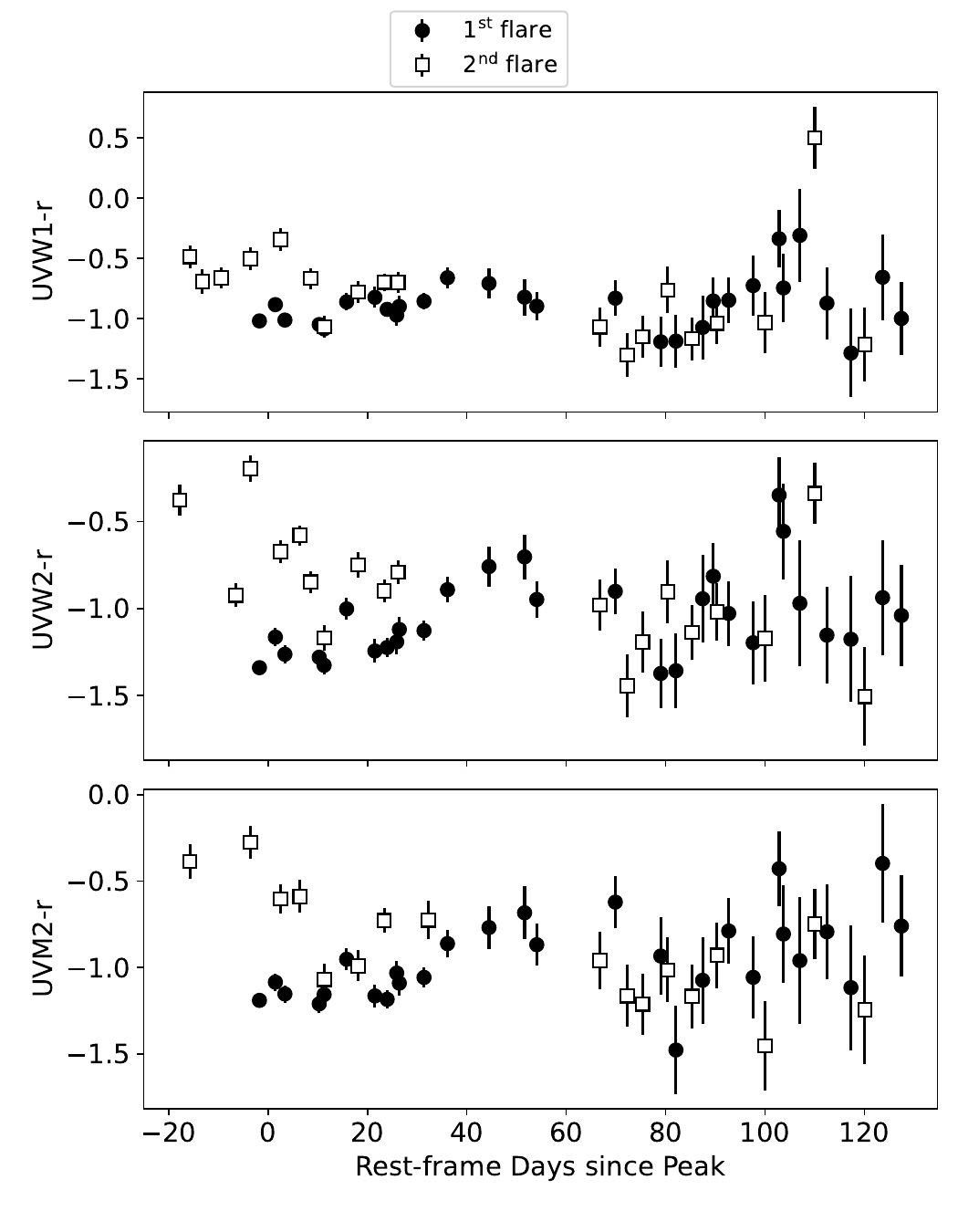}
\caption{The ultraviolet to optical color evolution of both \dbl\ flares is roughly constant in time, as seen in other optical-ultraviolet TDEs. However, the small but significant differences in color evolution between the flares from peak luminosity to late times is enough to argue against both flares being two lensed images of the same underlying event.
\label{EDfig:colors}}
\end{figure}

We perform two tests to check whether the two flares of \dbl\ could be two lensed images of the same TDE. First, we measure the ultraviolet to optical color evolution of both flares of \dbl\ (Fig. \ref{EDfig:colors}), which shows that both flares evolve slightly differently. In a lensing scenario, different colors between images can be explained by different amounts of dust extinction along the different light paths. However, the color differences should be constant in time, which is not the case for \dbl. The slight difference in color evolution between the flares thus already strongly disfavors a lensing scenario for \dbl.

Second, we examine whether the two flares can be two time-delayed images of a single lensed TDE given the timescale between the two flares, their redshift, and peak brightness ratio. Since both events have spectra that place them unambiguously at the redshift of the host, $z=0.0284$, we primarily consider if there may be an intervening mass between us and the host to account for lensing. In what follows, we assume a point mass lens. Other common alternatives (for example, a Singular Isothermal Sphere lens) should lead to similar conclusions. The deflection angle at an angular distance $\theta$ from a point mass of mass $M$, is given by $\alpha(\theta)=\frac{4GM}{c^{2}\theta} \frac{D_{\rm ls}}{D_{\rm l}D_{\rm s}}$, where $G$ is the gravitational constant and $c$ the speed of light. $D_{\rm l}$, $D_{\rm s}$, $D_{\rm ls}$ are the angular diameter distances to the lens (located at $z_{l}$), to the source (located at $z_{s}$), and between the lens and the source, respectively. A point mass lens would generally lead to two images appearing, one on each side of the lens. 
The magnification of the two images in this case is given by \citep{Schneider1992}:
\begin{equation}
\mu_\pm = \frac{u^{2}+2}{2u\sqrt{u^{2}+4}}\pm\frac{1}{2} ,
\label{mags}
\end{equation}
where $u$ is the angular separation of the source from the point mass in units of the Einstein angle, $u\equiv\beta\theta_{\rm E}^{-1}$. The Einstein angle  $\theta_{\rm E} = \left(\frac{4GM}{c^2} \frac{D_{\rm ls}}{D_{\rm l}D_{\rm s}}\right)^{1/2}$ is the radial position of the ring that would form around a point (or circularly symmetric) lens were the source lying behind its center along the line of sight to it. The time delay between the two images in this case can be written as \citep{Schneider1992}:
\begin{equation}
\Delta t=  \frac{4GM}{c^{3}} (1+z_{\rm l}) \tau(u) ,
\label{eqTDScheinder}
\end{equation}
where
\begin{equation}
  \tau(u)=\frac{1}{2}u\sqrt{u^{2}+4} + \ln\frac{\sqrt{u^{2}+4} +u}{\sqrt{u^{2}+4} -u}  .
\label{eqTDScheinder2}
\end{equation}

For the following calculations we assume a standard $\Lambda$CDM cosmology with $H_0=70$ km~s$^{-1}$~Mpc$^{-1}$, $\Omega_{\Lambda}=0.7$, and $\Omega_\mathrm{m}=0.3$. We adopt an angular diameter distance of $D_{\rm s}=1.1751\times10^{8}$ pc to the source, which corresponds to $z=0.0284$, and place a fiducial lens at $z=0.01$, yielding $D_{\rm l}=4.2307\times10^{7}$ pc. Using the above equations we find that such an intervening point-mass lens should be as massive as $\sim10^{13}$ M$_{\odot}$ to explain the time delay ($\sim$700 days) and magnification ratio ($\sim$2:1) of the two events, were they counter images of the same lensed event. Given the low redshift of the host and the lack of any optical signature of lensing or of an intervening galaxy, it is not likely that such a lensing mass is found exactly between us and the host galaxy. This result does not strongly depend on the redshift of the fiducial intervening lens, and a similarly large mass is required even if the point lens is assumed to lie within the host galaxy. Similarly, any reasonably massive point lens within our galaxy is also not capable in the above framework to generate images with sufficient time delay as observed. 

We conclude that it is very unlikely that the two events are time delayed images of the same lensed event, but are instead most likely two separate events.

\subsubsection{Two Unrelated TDEs}\label{sec:two_unrelated_tdes}

As stated above, the host-galaxy of \dbl\ is a quiescent Balmer-strong galaxy. Such galaxies have been shown to have an enhanced TDE rate \citep{Arcavi2014, French2016}. More specifically, TDE rates in such galaxies correlate strongly with their Lick H$\delta_A$ index. We use the 15 optical TDEs with broad lines in their spectra (i.e. the same class of events as \dbl) from the \cite{French2020SSRv} TDE host galaxy sample to estimate the probability of seeing two unrelated TDEs in a galaxy with similar Lick H$\delta_A$ index as that of \dbl\ within 700 days of each other. 

We obtain Lick H$\delta_A$ indices from the MPA-JHU catalogues of absorption line indices and emission line flux measurements \citep{Kauffmann2003,Brinchmann2004,Tremonti2004} retrieved through the SDSS III Science Archive Server. We calculate the fraction of TDEs in three H$\delta_A$ bins (equally spaced between 1.3 and 6\,\AA\, with the host of \dbl\ belonging to the first, i.e. lowest H$\delta_A$, bin) relative to the fraction of galaxies in each bin. This gives us a TDE rate enhancement for each galaxy bin of $\times$10.09, $\times$12.23, and $\times$132.43 (from the lowest H$\delta_A$ bin to the highest, respectively). We then multiply this enhancement factor by a global TDE rate between $10^{-5}$ and $10^{-4}$ events per galaxy per year \citep{Stone2020}. This gives us the TDE rate for each H$\delta_A$ bin, which we use, assuming Poisson statistics, to calculate the probability for two events to occur within 700 days of each other in that bin. We then multiply this number by the total number of TDEs discovered to date in that galaxy bin. We assume that 50 optical-ultraviolet TDEs have been discovered to date in all galaxy types, and that they are distributed between galaxy types in the same way as in the \cite{French2020SSRv} sample. 

We find that the probability of observing two unrelated TDEs within 700 days of each other in a galaxy sharing the same H$\delta_A$ bin as that of the \dbl\ host galaxy is 0.037--0.368\%, depending on the assumption of the global TDE rate ($10^{-5}$--$10^{-4}$ events per galaxy per year, respectively). For a recent global rate estimate of $3.2\times10^{-5}$ events per galaxy per year \citep{Yao2023}, the probability that the two flares of \dbl\ are of unrelated TDEs is 0.12\%. 

This result is not sensitive to the number of bins chosen (using 4 bins we find 0.14\%, and with 5 bins we find 0.16\%; the small sample of host galaxies precludes splitting the population to 6 bins as in that case, the bin containing the host of \dbl\ has zero galaxies in it, precluding a rate enhancement calculation for it). The small number statistics and uncertainty in the global TDE rate prohibit a precise chance coincidence calculation, but our results indicate that it is of order $\sim$0.01--0.1\%.

We repeat this analysis for AT\,2020vdq, a recently claimed repeating TDE \citep{Somalwar2023} for which the H$\delta_A$ index of the host galaxy \citep{Somalwar2023vlass} is 4.50$\pm$0.13\,\AA\ (J. Somalwar, private communication) and the time interval between the flares is 2.6 years. We find that the probability of observing two unrelated TDEs for these conditions is 0.86--8.61\%, depending on the assumption of the global TDE rate ($10^{-5}$--$10^{-4}$ events per galaxy per year, respectively). This is more than an order of magnitude larger than for \dbl. Thus, even if both flares of AT\,2020vdq were TDEs, it is much less clear than for \dbl\ that these two TDEs are related to the disruption of the same star. We conclude that \dbl\ is a much more likely repeating TDE than AT\,2020vdq.

\begin{figure*}
    \centering
    \includegraphics[width=0.8\textwidth]{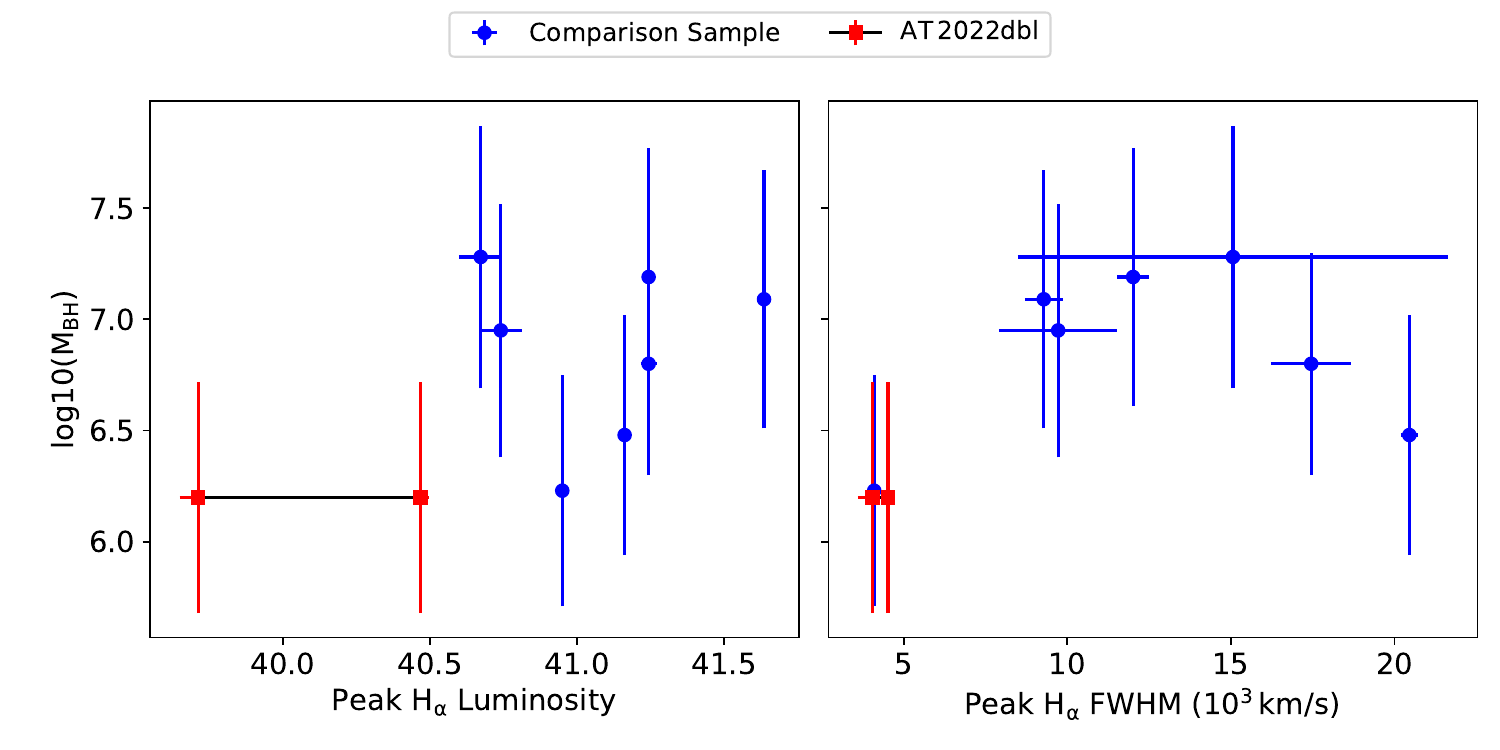}
    \caption{Black hole mass versus TDE $\halpha$ luminosity (left) and FWHM (right) around peak bolometric luminosity for \dbl\ (red) and the comparison sample in Figure \ref{EDfig:halpha} (blue). We do not find any strong correlations between black hole mass and $\halpha$ luminosity or width, indicating that the similarity of \halpha\ properties of the two \dbl\ flares is not solely related to the SMBH, but related to the star and/or its orbit. This strengthens the claim that the two flares of \dbl\ are related to the same star.\label{EDfig:ha_smbh}}
\end{figure*}

In addition, we argue that the likelihood of both flares of \dbl\ being of unrelated TDEs is even lower considering the similarity of the \halpha\ properties of both flares, compared to the spread of properties in a comparison sample of H-rich TDEs (Figure \ref{EDfig:halpha}). For two unrelated stars to produce the same \halpha\ properties would imply that these properties are set mainly by the black-hole rather than by the disrupted stars or their orbits. While it is not know what sets the properties of \halpha\ lines in TDEs, if they were driven primarily by the black hole, we would expect to see a correlation between \halpha\ emission properties and black-hole properties across TDEs. However, this is not seen. The presence of \halpha\ in a TDE spectrum does not correlate with black-hole mass \citep{Nicholl2022}, and when \halpha\ is present, neither do its width \citep{Panos2023} or luminosity (Figure \ref{EDfig:ha_smbh}). This suggests that at least the \halpha\ characteristics of a TDE must be influenced by its stellar and/or orbital properties, and hence that such similar spectra are difficult to explain as coming from two unrelated disruptions. Together with the rate argument above, we conclude that both flares of \dbl\ must relate to the same TDE.

\subsubsection{A Single TDE Around a SMBH Binary}

It has been suggested that post-starburst TDE host galaxies could be post-merger galaxies, harboring a binary SMBH in their center. However, while a binary black hole can significantly increase the TDE rate \citep{Chen2009,Wegg2011}, such systems have been disfavored as the dominant drivers of enhanced TDE rates in post-starburst galaxies due to the timescales involved \citep{French2017,Stone2019}. In addition, for \dbl\ to be caused by a SMBH binary, the orbital period of the SMBH binary would have to be comparable to 700 days, as this is the only timescale over which the presence of the secondary black hole would significantly modify the dynamics of the debris \citep[the debris exiting the Hill sphere of the primary yield a comparable estimate;][]{Coughlin2018}. In this case, a total black hole mass of $10^6\Msun$ implies a binary semi-major axis which is well within the sphere of influence of each black hole (since the semi-major axis scales as $\mbh^{1/3}$ for a given period, changes in the total mass at the order of magnitude level do not result in significant changes to the semi-major axis). While the cross section for tidal disruption is increased due to the larger geometrical area of the binary, at these small separations, the TDE rate would be enhanced by a factor of $\sim$2 at most, since such a binary would cause the majority of stars to be ejected prior to disruption \citep{Coughlin2017,Darbha2018}. Therefore, the double flare of \dbl\ does not necessarily favor binary black holes as the drivers of enhanced TDE rates in post-starburst galaxies.

Still, we check whether the second outburst of \dbl\ could be due to a companion black hole. While the secondary black hole does generate stochasticity in the accretion rate onto the primary, this is usually at the expense of producing a sudden drop in the accretion rate \citep{Ricarte2016}, which is not observed here. Furthermore, we would expect the second brightening to be much less energetic than the first, also not observed here, owing to the fact that the amount of mass supplied to the black hole (i.e., the fallback rate) is itself much smaller at later times.

Alternatively, it could be that the second flare was due to the accretion of material by the secondary black hole. Again, however, we would expect the flare amplitude (or at least the total energy liberated) to be much lower than in the first flare, because the ejecta from the disruption by the primary black hole would be highly geometrically extended and the total mass available to the secondary black hole considerably reduced. It is also difficult to see why the second accretion event would be spectroscopically nearly identical to the first, given that the geometry of the accretion flow and the accretion rate would be different. 

We conclude that the double flare of \dbl\ and its occurrence in a post-starburst galaxy do not constitute evidence for the presence of an SMBH binary.

\subsubsection{Emission From the Less-Bound Debris Tail}

A star on a parabolic orbit that is disrupted (partially or completely) produces two tidal tails of debris, one of which is bound while the other is unbound. The reason for this is that the tidal field imparts a spread in the specific Keplerian energy of 
\begin{equation}
    \Delta\epsilon \simeq \frac{GM_{*}}{R_{*}}\left(\frac{\mbh}{\mstar}\right)^{1/3}\label{deltaepsilon}
\end{equation}
(where \mstar\ is the mass of the disrupted star and \rstar\ its radius) to the debris as the star passes through the tidal disruption radius \citep{lacy82}, meaning that the most-unbound debris has a specific energy $+\Delta\epsilon$. Equation \eqref{deltaepsilon} relies on the assumption that the star is approximately hydrostatic as it passes through the tidal radius. This spread in energy has been verified by a number of numerical investigations when the star is described by a $5/3$-polytrope and is completely destroyed \citep{lodato09, GR2013,steinberg19,norman21}, but it is less clear that it should hold for partial disruptions or for different stellar structures.

When the star is bound to the black hole by an amount $\epsilon_{\rm c}$ (i.e., $\epsilon_{\rm c}$ is the specific energy of the surviving core), this energy barrier must be overcome to completely unbind material from the black hole, meaning that the most unbound (or ``least bound'' if the specific energy is ultimately negative) material has a specific energy $\epsilon = \epsilon_{\rm c}+\Delta\epsilon$.
This implies that the ratio of $\epsilon_{\rm c}/\Delta\epsilon$ is a measure of the importance of the specific energy of the core: if $\epsilon_{\rm c}/\Delta\epsilon \ll 1$, then the core binding energy is unimportant and the system is effectively parabolic (as concerns the ejection of mass on hyperbolic trajectories), whereas if $\epsilon_{\rm c}/\Delta \epsilon \gg 1$, the entire stream is bound to the black hole \citep{hayasaki18}. Adopting an orbital period of $700$ days and a black hole mass of $\mbh = 10^6\,\Msun$, the energy-period relationship of a Keplerian orbit gives $\epsilon_{\rm c} \simeq -15 G\mstar/\rstar$ for a solar-like star. Since $\Delta \epsilon = 100 G\mstar/\rstar$, we would expect at most $\sim 15\%$ of the ``unbound'' tail to be gravitationally bound to the black hole (this is an upper limit, as some fraction of the material will accrete back onto the star). Because the return time of the debris must be at least as long as the orbital period of the star, a very firm upper limit on the accretion rate onto the black hole from the less-bound tail is (again, for a solar-like star) $\sim 0.07\,\Msun/(700\rm{ d}) \simeq 0.05\,\Msun$ yr$^{-1}$. This is approximately two orders of magnitude below the peak from the main flare (see Figure \ref{fig:hydrosimulations}), implying that the second flare cannot -- barring extreme changes to the accretion efficiency of the black hole -- be due to the less-bound debris.

\subsubsection{The Repeated Disruption of the Same Star}

Having disfavored gravitational lensing of the same TDE, two unrelated TDEs, a single TDE around a SMBH binary, and the accretion from a debris tail; given the photometric and spectroscopic similarity of the two flares; and having presented analytical and numerical models that are consistent with the two flares arising from the disruption of the same star, we conclude that \dbl\ is most likely the repeated disruption of the same star, with at least the first disruption being partial. 

\subsection{Dynamical Scenario for the Creation of \dbl}

For a black hole of mass $\mbh\ \approx 10^6-10^{6.7}\,\Msun$, a 700-day orbit implies a semi-major axis $a$ of 2.3--4.0$\times10^{15}$\,cm. This is much smaller than the radius of influence of the black hole (which for the host galaxy of \dbl\ is $\sim$3--7$\times10^{18}$\,cm), from where most disrupted stars originate, and tidal dissipation alone cannot bind the star to the black hole sufficiently tightly to produce a 700-day period \citep{cufari22}. 

Instead, the star could have been captured through the Hills mechanism \citep{Hills1988}. According to this mechanism, the center of mass of a binary star system with mass $M_b$ and separation $a_b$ approaches the SMBH on an orbit with a pericenter distance smaller than $R_a=a_b\left(\frac{\mbh}{M_b}\right)^{1/3}$. At that distance from the SMBH, the binary is disrupted, with one star becoming unbound and ejected from the system, and the other star remaining bound to the black hole. The bound star is expected to have an orbital period $\lesssim$10 years (see below), consistent with the period measured here. 

For any mass to be removed from the star, the pericenter distance of its orbit $R_p$ should be $\lesssim$2 times the stellar tidal disruption radius \citep{coughlin22} $R_t=\rstar\left(\frac{\mbh}{\mstar}\right)^{1/3}$. While this is typically smaller than the binary tidal disruption radius $R_a$, the captured star could be put onto such an orbit directly by the Hills mechanism \citep{cufari22}, or it could shrink its orbit and increase its eccentricity after being captured through two-body scatterings (see below).

Another way to produce the 700-day orbit is through eccentric Kozai-Lidov oscillations \citep{Kozai1962,Lidov1962} from a SMBH binary, combined with two-body relaxation \citep{Melchor2024}. In this scenario, the second black hole would most likely be at separations of order the radius of influence from the disrupting black hole, and therefore would not affect the observed flare, only the dynamics leading up to disruption.

\subsubsection{The Orbital Period of \dbl}

The difference between observed light curve peaks $\Delta T_{\rm obs}$, assuming that the luminosity tracks the fallback rate, is $\Delta T_{\rm obs} = T_{\rm orb} - T_{\rm peak, 1} +T_{\rm peak, 2}$, where $T_{\rm peak, 1}$ and $T_{\rm peak, 2}$ are the times to reach the peak fallback rate following the first and second pericenter passage of the star, respectively. These two times will not be identical because the star is imparted rotation \citep{Ryu2020a,Bandopadhyay2024} and loses mass upon being partially disrupted. Hence the orbital period of the star can only be approximated from the time between peaks \citep[the observation of the sudden shutoff of emission, on the other hand, could directly constrain the time at which the star reaches pericenter;][]{Wevers2023}.

That being said, we now show that a period of $\sim$700 days is consistent with theoretical expectations from the Hills mechanism. The tidal radius of a binary star system with semimajor axis $a_b$ and total mass $M_b$ is:
\begin{equation}\label{eq:r_t_binary}
R_a = a_b\left(\frac{\mbh}{M_{b}}\right)^{1/3}
\end{equation}
As the binary is ripped apart, one of the stars is captured on an orbit that has a specific binding energy about the black hole of:
\begin{equation}
\epsilon_{*} = \frac{GM_{b}}{a_{b}}\left(\frac{\mbh}{M_{\rm b}}\right)^{1/3}
\label{epsstar}
\end{equation}
This then yields a period of the captured star of:
\begin{equation}\label{eq:t_star}
T_{*} = \frac{\pi}{\sqrt{2}}\frac{G\mbh}{\epsilon_{*}^{3/2}} \simeq \frac{a_{b}^{3/2}}{\sqrt{GM_b}}\left(\frac{\mbh}{M_b}\right)^{1/2}
\end{equation}
%This is the same expression that one recovers from a TDE but this is now in terms of the properties of the binary.
For the binary to avoid being disrupted before reaching $R_t$, we require \citep{quinlan1996}:
\begin{equation}\label{eq:max_a}
\sqrt{\frac{GM_{b}}{a_{b}}}\gtrsim \sigma \quad \Rightarrow \quad a_{b} \lesssim \frac{GM_{b}}{\sigma^2}
\end{equation}
where $\sigma$ is the velocity dispersion of the stars in the nucleus of the galaxy.
Combining this with Equation \ref{eq:t_star} yields:
\begin{equation}
T_{*} \lesssim \frac{GM_{b}}{\sigma^3}\left(\frac{\mbh}{M_{b}}\right)^{1/2}
\end{equation}
Adopting an \mbh-$\sigma$ relation of the form \citep{Merritt2000}:
\begin{equation}
\mbh = M_0\left(\frac{\sigma}{\sigma_0}\right)^{n} \quad \Rightarrow \quad \sigma  = \sigma_{0}\left(\frac{\mbh}{M_0}\right)^{1/n},
\end{equation}
the equation for $T_{*}$ becomes:
\begin{equation}
T_{*} \lesssim \frac{G\sqrt{M_{b}}}{\sigma_0^3}\left(\frac{\mbh}{M_0}\right)^{-3/n}\mbh^{1/2},
\end{equation}
which can be re-written as:
\begin{equation}
T_{*} \lesssim \frac{G\sqrt{M_{b}M_{0}}}{\sigma_0^3}\left(\frac{\mbh}{M_0}\right)^{\frac{n-6}{2n}}
\end{equation}
Recent estimates of the \mbh-$\sigma$ relation give $n = 5.1$, $M_0 = 1.9\times 10^{8} \Msun$, and $\sigma_0 = 200$ km\,s$^{-1}$ \citep{McConnell2011}. This then gives:
\begin{equation}\label{eq:max_period}
T_{*} \lesssim 7.2\left(\frac{M_{b}}{\Msun}\right)^{1/2}\left(\frac{\mbh}{M_0}\right)^{-3/34}\textrm{ yr}
\end{equation}
This suggests that there could be many objects with periods longer than the $\sim$700-day orbit of \dbl, and that the binary that produced the captured star was particularly hard, i.e., with a semimajor axis significantly smaller than $\sim G M_{b}/\sigma^2$. In addition, Equation \eqref{epsstar} is a most-likely value, and while binding energies larger than this are rare, there could be a substantial number \citep{cufari22} of captured objects with periods that are even longer than the limit in Equation \ref{eq:max_period}.

Changes to the orbital period could be achieved by imparting energy -- either through rotation or internal/oscillatory modes -- to the surviving core. This effect is important when the imparted energy is comparable to (or at least not too dissimilar from) the binding energy of the core to the black hole. In terms of the orbital period $T$, the latter is $\epsilon_{\rm c} = \left(2 \pi G \mbh/T)\right)^{2/3} / 2$, which for $\mbh = 10^6\,\Msun$ and $T = 700$ days is $\epsilon_{\rm c} \simeq 3\times 10^{16}$ erg g$^{-1}$. The energy imparted via tides can at most be equal to the binding energy of the star, meaning that the change in the specific energy satisfies $\Delta \epsilon \le \epsilon_{*} \simeq GM_{*}/R_{*} \simeq 2\times 10^{15}$ erg g$^{-1}$, where we adopted a mass-radius relationship $R_{*} = R_{\odot}\left(M_{*}/M_{\odot}\right)$ for a main sequence star \citep{demircan91}. The ratio of the energy imparted via tides to the binding energy of the orbit is therefore $\lesssim 1/15$, but this is likely a substantial over-estimate, given that the mass powering the accretion is a very small fraction of a solar mass. In the limit where the star is effectively at the partial disruption radius and a low amount of mass stripped, the change in the binding energy of the star is \citep{cufari23} $\Delta \epsilon \simeq 1\%\epsilon_{*}$, and we would therefore expect the relative change in the period to be $\Delta T/T \simeq -3/2\times \Delta\epsilon/\epsilon_{\rm c} \simeq 0.1\%$, implying an absolute change in the orbital period of $\Delta T \lesssim 1$ day. 

\subsubsection{Formation of \dbl\ Through the Hills Mechanism and Angular Momentum Relaxation}

With an orbital period of $T_{*} \approx 700$ days, and a pericenter distance $R_{p}$ comparable to the tidal disruption radius $R_{t}$, the orbital eccentricity is constrained to:
\begin{equation} \label{eq:ecc_current}
\begin{split}
    1 - e& \approx \beta_*^{-1} (2\pi)^{2/3}  \frac{R_*}{(GM_* T_{*}^2)^{1/3}}\\& \approx 0.006 \left( \frac{T_{*}}{700 \, \rm d} \right)^{-2/3} \left( \frac{R_*}{R_\odot} \right) \left( \frac{M_*}{M_\odot} \right)^{-1/3} \left( \frac{\beta_*}{0.5} \right)^{-1}
    \end{split}
\end{equation}
where we normalize to $\beta_* \equiv R_{t}/R_{p} = 0.5$ as the penetration factor appropriate for a partial disruption. On the other hand, stars in a binary of mass $M_{b}$ captured via the Hills mechanism typically have an eccentricity of approximately \citep{cufari22}:
\begin{equation} \label{eq:ecc_Hills}
\begin{split}
    1-e_{\rm Hills} &\approx \frac{2}{\beta_{b}} (M_{b}/{\mbh})^{1/3} \\&\approx 0.025 \, \beta_{b}^{-1} \, \left( \frac{\mbh}{10^6 \, \rm M_\odot} \right)^{-1/3} \left( \frac{M_{\rm b}}{2 \, \rm M_\odot}\right)^{1/3}
    \end{split}
\end{equation}
where here we normalize to a penetration factor for the binary $\beta_{b} \equiv R_{a}/R_p = 1$ (here $R_p$ is the pericenter distance of the binary from the black hole), as the tidal splitting of binaries in galactic nuclei is typically governed by an empty loss-cone regime \citep{Stone2020}. 

Thus, either the binary penetration factor $\beta_b$ is roughly 4, or the orbital eccentricity of the captured star needs to be excited from its initial value to its value inferred at disruption (Eq.~\ref{eq:ecc_current}). This can happen through two-body scatterings with other field stars, a process that would occur over the angular momentum relaxation time of approximately \citep{Merritt_2013}:
\begin{equation}
\begin{split}
    \tau_{\rm 2B}^{J} &\approx \frac{P_{\rm orb}}{2\pi N_*(R_a)} \left( \frac{\mbh}{M_*} \right)^2 (\ln{\Lambda})^{-1} (1-e_{\rm Hills}) \\&\approx 5\times 10^{6} \, {\rm yr} \, \left( \frac{P_{\rm orb}}{700 \, \rm d} \right)^{1/6} \left( \frac{\ln \Lambda}{10} \right)^{-1} \left( \frac{\mbh}{10^6 \, \rm M_\odot} \right)^{0.18}
\end{split}
\end{equation}
where $N_*(R_a)$ is the number of stars of semi-major axis comparable to $R_a$ and $\ln{\Lambda} \approx \ln{\mbh/M_*} \approx 10$ is the Coulomb logarithm. Here we assume a Bahcall-Wolf density profile \citep{BW_77}, such that $N_*(R_a) \approx (\mbh/M_*) (R_a/G\mbh/\sigma^2)^{5/4}$, and the $\mbh-\sigma$ relation of \cite{KH2013}. For the orbital parameters relevant here, this time is shorter than the circularization time due to gravitational waves, which is given by \citep{Peters1964}:
\begin{equation}
\begin{split}
    \tau_{\rm GW}^{E} &\approx \frac{12 \sqrt{2}}{85} \frac{G\mbh}{c^3} \frac{\mbh}{M_*} \left( \frac{a}{G\mbh/c^2}\right)^4 (1-e_{\rm Hills})^{7/2} \\&\approx 4\times 10^9 \, {\rm yr} \; \beta_{b}^{-7/2} \left( \frac{\mbh}{10^6 \, \rm M_\odot} \right)^{-11/6} \left( \frac{P_{\rm orb}}{700 \, \rm d} \right)^{8/3}
\end{split}
\end{equation}

\begin{figure}
    \centering
    \includegraphics[width=\linewidth]{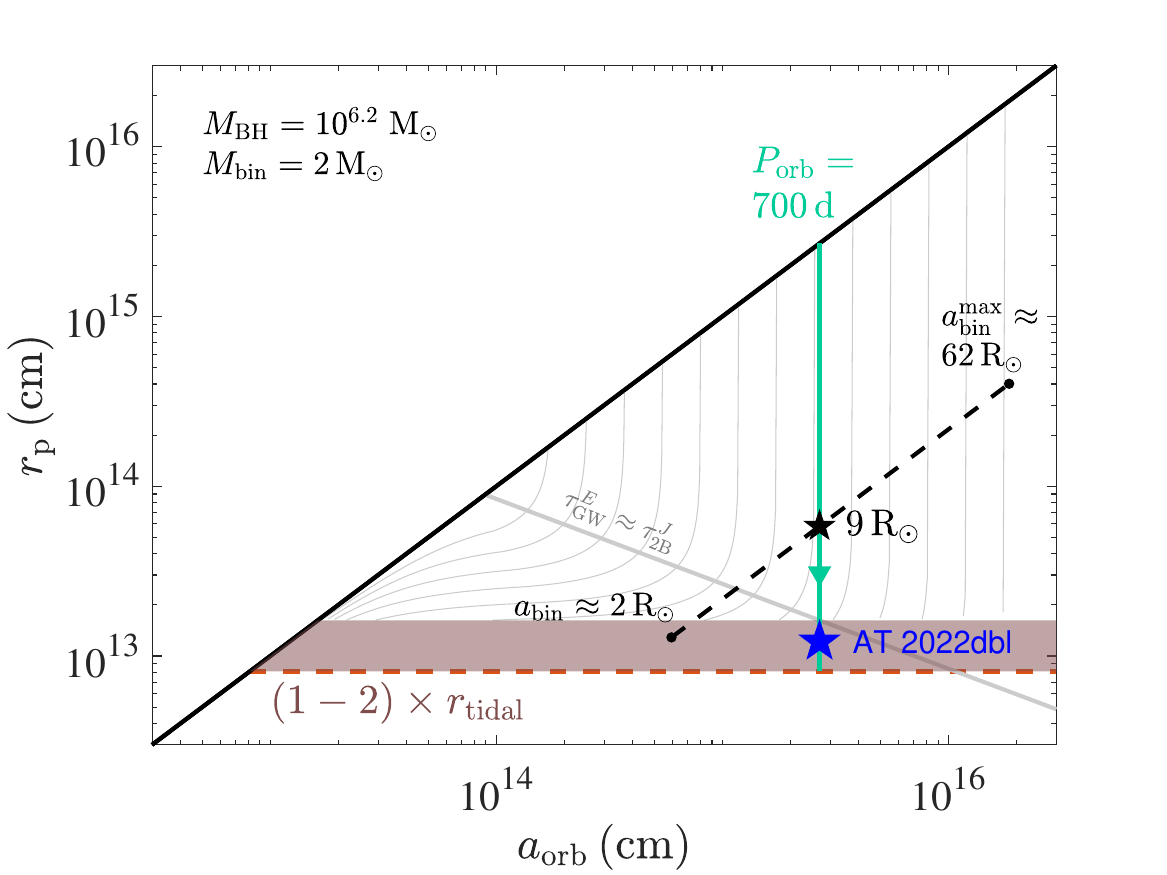}
    \caption{Illustration of a possible Hills-mechanism formation channel for \dbl. Orbits of a binary star system of total mass $M_{\rm bin}=2\,\rm M_\odot$ near a black hole of mass $M_{\rm BH}=10^{6.2} \, \rm M_\odot$ are shown in a phase space of pericenter distance ($r_{\rm p}$) versus semi-major axis ($a_{\rm orb}$). The solid black diagonal line at $r_{\rm p}=a_{\rm orb}$ corresponds to circular orbits ($e=0$), while eccentric orbits lie below this line. The vertical green line denotes the approximate semi-major axis of \dbl, the shaded red region its pericenter distance (see text for details), and the blue star denotes their intersection. The black dashed diagonal line corresponds to the typical orbits of stars captured by the Hills mechanism, spanned by the range of possible binary separations $a_{\rm bin}$, from contact binaries ($a_{\rm bin}\approx 2\, \rm R_\odot$) to binaries on the soft/hard boundary given by the galactic velocity dispersion ($a_{\rm bin}\approx 62 \, \rm R_\odot$). A binary of intermediate separation, $a_{\rm bin} \approx 9 \, \rm R_\odot$, marked by the black star, would result in an orbital period similar to that of \dbl, yet with $r_{\rm p} \approx 10$ times the tidal radius. Angular momentum relaxation through two-body scatterings could excite the captured star's eccentricity without undergoing substantial changes in energy (and hence in $a_{\rm orb}$), bringing it towards the tidal radius on a nearly vertical track in this phase space (denoted by the green arrow). Thin grey lines show steady-state trajectories in this phase space, due to the combined effect of two-body relaxation and gravitational wave inspiral. Well above the thick gray line, orbits evolve primarily due to relaxation, along mostly vertical tracks, while below this line, gravitational wave evolution is more rapid, and orbits follow a nearly horizontal trajectory.\label{EDfig:hills}}
\end{figure}

Thus, the orbit will not undergo substantial changes in energy and period during its angular momentum relaxation to higher eccentricity. The hierarchy $\tau_{\rm 2B}^J < \tau_{\rm GW}^E$ is maintained even as $e$ increases to the (partial) disruption of the star, since $\tau_{\rm 2B}^J/\tau_{\rm GW}^E \propto (1-e)^{-5/2}$, such that as $(1-e)$ decreases by a factor of $\sim$4, the timescale ratio increases by a factor of $\sim$32. This process is illustrated in Figure \ref{EDfig:hills}.

In summary, assuming a Sun-like star captured from a binary system with mass ratio $\sim$1 by the Hills mechanism around a $10^6\,\Msun$ black hole, the following orbital parameters can reproduce a system like AT\,2022dbl. For a velocity dispersion $\sigma=200$ km\,s$^{-1}$ in the galaxy nucleus, the maximum allowed semi-major axis for the binary to survive down to the black hole is (Eq. \ref{eq:max_a}) $a_b\lesssim6.7\times10^{11}$ cm. The tidal radius for disrupting such a binary is then (Eq. \ref{eq:r_t_binary}) $\lesssim5.3\times10^{13}$ cm. The captured star would then have entered an orbit at a pericenter distance of $1.4\times10^{13}$ cm which is twice its tidal disruption radius $R_t=R_*\left(\mbh/M_*\right)^{1/3}=7\times10^{12}$ cm, to undergo a partial disruption with penetration factor $\beta_*=0.5$. This would entail an eccentricity of $e=0.993$ (Eq. \ref{eq:ecc_current}), making the semi-major axis $a=R_p/\left(1-e\right)=2.3\times10^{15}$ cm, corresponding to period of $P=2{\pi}a^{3/2}/\sqrt{G\mbh}=700$ days, as observed.

The formation of such systems to begin with is governed by the scattering rate of binaries in galactic nuclei onto highly eccentric orbits, leading to their tidal split-up. If binaries of the relevant range of orbital separations constitute a fraction $f_{\rm b}$ of all stars within the black hole's sphere of influence, the formation rate of such systems will be approximately $f_{\rm b} \mathcal{R}_{\rm TDE}$, where $\mathcal{R}_{\rm TDE}$ is the overall TDE rate. Systematic classification of full vs. repeating TDEs will thus be useful in constraining the dynamics and demographics of stellar populations within galactic nuclei.

\subsection{Implications}

The similarity of \dbl\ to the class of optical-ultraviolet TDEs in its total energy, temperature, spectral features and host galaxy characteristics raises the question of whether all members of the class are in fact partial disruptions \citep[as tentatively suggested also by][]{Liu2024}.

The existence of multiple flares of similar radiated energy in a seemingly normal TDE has clear implications for the missing energy problem in optical-ultraviolet TDEs. \dbl\ suggests that the total radiated energy in other TDEs may currently be underestimated given the uncertain recurrence times and number of flares associated with a given optical-ultraviolet TDE. 

Alternatively, it could be that both full and partial disruptions emit almost the same amount of energy in radiation with nearly identical emission properties. This could happen if the luminosity of the accretion flow is roughly limited by the Eddington luminosity, even if the accretion rate is not. Indeed, depending on the black hole mass assumed for \dbl, the first flare could be Eddington limited\footnote{Using the host-galaxy black hole estimate we find a peak Eddington ratio of $0.47^{+1.11}_{-0.32}$ for the first flare. The lower black hole mass estimates from TDEMass imply and even higher ratio.}. The luminosities of other TDEs are also comparable to the Eddington luminosity for the more common low-mass black holes \citep{Yao2023} and the theoretically predicted fallback rates are super Eddington \citep{Rees1988}. If this is the case it can solve the missing energy problem through super-Eddington accretion in which the energy goes into the black hole and into outflows, rather than into radiation. In this interpretation, the distinguishing feature between partial and complete disruption would primarily be the post-peak decline rate (being steeper for partial disruptions), which is consistent with both the observations presented here and theoretical predictions \citep{Rees1988,Phinney1989,GR2013,CN2019,Krolik2020,Bandopadhyay2024}. 

Indeed, both flares have similar light curves, with the second being fainter and having a shallower post-peak decline (${\sim}t^{-5/3}$) than the first (${\sim}t^{-2.7}$). However, it is not possible to determine whether the second flare continued to decline at this rate or whether it simply had a broader peak, and its decline rate later steepened to resemble that of the first flare. 

This leads to two possibilities regarding the fate of the star and future flares of \dbl: either both disruptions were partial, or the first was partial and the second was the full disruption of the star. The existence of a third flare (possibly very similar to the first two; top panel of Figure \ref{fig:hydrosimulations}) in early 2026 will determine whether the second flare was also due to a partial disruption, or whether the second flare could have been the full disruption of the star. 

The first possibility implies that all optical-ultraviolet TDEs, regardless of initial luminosity decline rate, could be partial disruptions, while the second possibility implies that only some optical-ultraviolet TDEs are partial disruptions, exhibiting similar observational properties as full disruptions. Whichever turns out to be the correct interpretation, it will have far reaching implications for our understanding of TDEs since most models to date assume that all optical-ultraviolet TDEs are full disruptions (an assumption ruled out here by either interpretation). 

\section{Summary and Conclusions}

We presented and analyzed observations of the double-flare TDE \dbl. The spectral features, blackbody temperature, luminosity and total integrated energy of each flare of \dbl\ are typical of the optical-ultraviolet TDE class \citep{VV2020,Hammerstein2023}, as are the radio detections from shortly after optical discovery \citep[indicating a sub-relativistic outflow with no luminous delayed radio flare in the first two years;][]{Alexander2016, horesh_2021a}. The host galaxy of \dbl\ is a quescent Balmer-strong galaxy, which is also typical of optical-ultraviolet TDE hosts \citep{Arcavi2014,French2020SSRv} and was even pre-selected four years before the discovery of \dbl\ as a likely TDE host \citep{French2018}. 

Thus, all of the characteristics of \dbl\ place it in the class of optical-ultraviolet TDEs, without being an outlier in any parameter, except for having a second flare approximately 700 days after the first flare. We show that the second flare exhibits a very similar light curve (albeit fainter and with a slower post-peak decline in the ultraviolet bands) and nearly identical spectra as the first flare. 

Given the time separation and slight differences in ultraviolet to optical color evolution between the two flares, we are able to rule out that the two flares are due to gravitational lensing of a single TDE. Our analysis of the host-galaxy of \dbl\ strongly disfavors two unrelated TDEs as the origin of the two flares, and the light curves of both flares disfavor the disruption of the same star by two different SMBHs. We presented both analytical and numerical models that are consistent with the scenario of both flares being the disruption of the same star by the same SMBH, with at least the first flare being due to a partial disruption of the star. We have further presented a possible dynamical scenario that can reproduce the disruption parameters.

We conclude that \dbl\ represents two consecutive disruptions of the same star, which is on a bound orbit (with an orbital period of approximately 700 days) about a SMBH, and for which at least the first flare was the result of a partial disruption. Given that until now, candidate repeating partial disruptions were either for events with spectra and light curves not typical of the optical-ultraviolet TDE class, for events for which optical spectra or photometry were not taken during one of the flares, or for events in host galaxies capable of producing multiple TDEs on the observed timescales, we determine that \dbl\ is the first robust case of a partial disruption appearing as an otherwise ``normal'' optical-ultraviolet TDE.

As such, this has far-reaching implications for the class of optical-ultraviolet TDEs as a whole. It is possible that other members of this class (or even all of them) are also partial disruptions, just on longer return timescales (or of stars on single passes). If this is the case, it requires a re-assessment of the emission mechanisms, rates and processes driving the host-galaxy preference of optical-ultraviolet TDEs.\\
~\\
This research was supported in part by grant NSF PHY-2309135 to the Kavli Institute for Theoretical Physics (KITP), where part of this work was done, during the ``Towards a Physical Understanding of Tidal Disruption Events'' program. We are grateful to Ashish Meena for helpful discussion, and to the anonymous referee for useful comments. 

LM acknowledges support through a UK Research and Innovation Future Leaders Fellowship (grant number MR/T044136/1). Support for LM at Tel Aviv University was provided by the European Research Council (ERC) under the European Union’s Horizon 2020 research and innovation program (grant agreement number 852097).
IA acknowledges support from the ERC under the European Union’s Horizon 2020 research and innovation program (grant agreement number 852097), from the Israel Science Foundation (ISF; grant number 2752/19), from the United States - Israel Binational Science Foundation (BSF; grant number 2018166), and from the Pazy foundation (grant number 216312).
MN, MA, JF, DAH, CM, EPG and GT are supported by the United States National Science Foundation (NSF) grants AST-1911225, and AST-1911151.
AB and ERC acknowledge support from the National Aeronautics and Space Administration (NASA) through the Neil Gehrels Swift Guest Investigator Program (proposal number 1922148). Additional support for AB was provided by NASA through Chandra Award Number 25700383 issued by the Chandra X-ray Observatory Center, which is operated by the Smithsonian Astrophysical Observatory for and on behalf of NASA under contract NAS8-03060, and by NASA through the FINESST program, grant 80NSSC24K1548. ERC acknowledges additional support from the NSF through grant AST-2006684 and from NASA through the Astrophysics Theory Program (grant number 80NSSC24K0897).
IL acknowledges support from a Rothschild Fellowship, The Gruber Foundation and a Simons Investigator grant (number 827103).
AH is grateful for support by the BSF (grant number 2020203), the ISF (grant number 1679/23), and by the Sir Zelman Cowen Universities Fund. 
CJN acknowledges support from the Science and Technology Facilities Council (grant number ST/Y000544/1) and the Leverhulme Trust (grant number RPG-2021-380).
KDF acknowledges support from NSF grant AST–2206164.
AZ acknowledges support by the BSF (grant number 2020750), the NSF (grant number 2109066), the Ministry of Science and Technology, Israel, and by the ISF (grant number 864/23).
PC acknowledges support via the Research Council of Finland (grant number 340613).
MJD is funded by the UK Science and Technology Facilities Council (STFC) as part of the Gravitational-wave Optical Transient Observer (GOTO) project (grant number ST/V000853/1).
JL, MP and DO acknowledge support from a UK Research and Innovation Fellowship (MR/T020784/1). 

This work makes use of data from the Las Cumbres Observatory global telescope network. 
This paper includes data collected by the TESS mission, which are publicly available from the Mikulski Archive for Space Telescopes (MAST). Funding for the TESS mission is provided by NASA’s Science Mission directorate. 
We acknowledge the staff who operate and run the AMI-LA telescope at Lord’s Bridge, Cambridge, for the AMI-LA radio data. AMI is supported by the Universities of Cambridge and Oxford, and by the European Research Council under grant ERC-2012-StG-307215 LODESTONE.
We thank the National Radio Astronomy Observatory (NRAO) for carrying out the Karl G. Jansky Very Large Array (VLA) observations.

\bibliography{2022dbl}{}
\bibliographystyle{aasjournal}

\end{document}

%% file: symbols.tex
\usepackage{float}
\usepackage{bm}

\newcommand{\todo}{\ifmmode \text{\color{red}{\Huge(\bullet)}} \else {\color{red}{\Huge$\bullet$}}\fi}
\newcommand{\tido}{\ifmmode {{\color{red}\bullet}} \else {\color{red}{$\bullet$}}\fi}

\newcommand{\E        }[1]{\ifmmode 10^{#1} \else $10^{#1}$\fi}
\newcommand{\tE        }[1]{\ifmmode \times10^{#1} \else $\times10^{#1}$\fi}
\newcommand{\til}{\ifmmode \sim \else $\sim$\fi}
\renewcommand{\~} {\ifmmode \sim \else $\sim$\fi}

\newcommand{\pc}	{\ifmmode \textrm{pc} \else pc\fi}
\newcommand{\kpc}	{\ifmmode \textrm{kpc} \else kpc\fi}
\newcommand{\ld}	{\ifmmode \textrm{l.d.} \else l.d.\fi}
\newcommand{\kms}	{\ifmmode \textrm{km\,s}^{-1} \else km\,s$^{-1}$\fi}
\newcommand{\cc}	{\ifmmode \textrm{cm}^{-3}    \else cm$^{-3}$\fi}
\newcommand{\cmii}	{\ifmmode \textrm{cm}^{-2}    \else cm$^{-2}$\fi}
\newcommand{\ergs}	{\ifmmode \textrm{erg\,s}^{-1} \else erg s$^{-1}$\fi}
\newcommand{\ergcms}	{\ifmmode \textrm{erg\,cm}^{-2}\,\textrm{s}^{-1} \else erg\,cm$^{-2}$\,s$^{-1}$\fi}
\newcommand{\ergcmsA}	{\ifmmode \textrm{erg\,cm}^{-2}\,\textrm{s}^{-1}\,\textrm{\AA}^{-1}
\else erg\,cm$^{-2}$\,s$^{-1}$\,\AA$^{-1}$\fi}
\newcommand{  \ergcmsHz  }{\ifmmode\textrm{erg\,cm}^{-2}\,\textrm{s}^{-1}\,\textrm{Hz}^{-1}
                       \else ergs\,cm$^{-2}$\,s$^{-1}$\,Hz$^{-1}$\fi}
\newcommand{\kev}	{\ifmmode \textrm{keV} \else keV\fi}

\newcommand{\mic}	{\ifmmode \textrm{\mu m} \else $\mu$m\fi}
\newcommand{\vFWHM}	{\ifmmode v_{\mbox{\tiny FWHM}} \else $v_{\mbox{\tiny FWHM}}$\fi}
\newcommand{\vBLR}	{\ifmmode v_{\mbox{\tiny BLR}} \else $v_{\mbox{\tiny BLR}}$\fi}
\newcommand{\sigBLR}	{\ifmmode \sigma_{\mbox{\tiny BLR}} \else $\sigma_{\mbox{\tiny BLR}}$\fi}
\newcommand{\vNLR}	{\ifmmode v_{\mbox{\tiny NLR}} \else $v_{\mbox{\tiny NLR}}$\fi}
\newcommand{\tauBLR}	{\ifmmode \tau_{\mbox{\tiny BLR}} \else $\tau_{\mbox{\tiny BLR}}$\fi}

\newcommand{\Hubble}	{\ifmmode \textrm{km\,s}^{-1}\,\textrm{Mpc}^{-1} \else km\,s$^{-1}$\,Mpc$^{-1}$\fi}
\newcommand{\NDunit}	{\ifmmode \textrm{Mpc}^{-3} \else Mpc$^{-3}$\fi}
\newcommand{\LFunit}	{\ifmmode \textrm{Mpc}^{-3}\,\textrm{mag}^{-1} \else Mpc$^{-3}$\,mag$^{-1}$\fi}
\newcommand{\MFunit}	{\ifmmode \textrm{Mpc}^{-3}\,\textrm{dex}^{-1} \else Mpc$^{-3}$\,dex$^{-1}$\fi}

\newcommand{\Msun}{\ifmmode M_{\odot} \else $M_{\odot}$\fi}
\newcommand{\Lsun}{\ifmmode L_{\odot} \else $L_{\odot}$\fi}
\newcommand{\Zsun}{\ifmmode Z_{\odot} \else $Z_{\odot}$\fi}
\newcommand{\mpyr}{\ifmmode \Msun\,\textrm{yr}^{-1} \else $\Msun\,\textrm{yr}^{-1}$\fi}

\newcommand{\qnote}{\ifmmode q_{0} \else $q_{0}$\fi}
\newcommand{\Hnote}{\ifmmode H_{0} \else $H_{0}$\fi}
\newcommand{\hnote}{\ifmmode h_{0} \else $h_{0}$\fi}
\newcommand{\anote}{\ifmmode a_{0} \else $a_{0}$\fi}
\newcommand{\tnote}{\ifmmode t_{0} \else $t_{0}$\fi}

\newcommand{  \Halpha   }{\ifmmode \textrm{H}\alpha \else H$\alpha$\fi}
\newcommand{  \halpha   }{\Halpha}
\newcommand{  \ha       }{\Halpha}
\newcommand{  \Hbeta    }{\ifmmode \textrm{H}\beta \else H$\beta$\fi}
\newcommand{  \hbeta    }{\Hbeta}
\newcommand{  \hb       }{\Hbeta}
\newcommand{  \Hgamma   }{\ifmmode \textrm{H}\gamma \else H$\gamma$\fi}
\newcommand{  \Hdelta   }{\ifmmode \textrm{H}\delta \else H$\delta$\fi}
\newcommand{  \Lya      }{\ifmmode \textrm{Ly}\alpha \else Ly$\alpha$\fi}
\newcommand{  \Lyb      }{\ifmmode \textrm{Ly}\beta \else Ly$\beta$\fi}
\newcommand{  \Pa       }{\ifmmode \textrm{P}\alpha \else P$\alpha$\fi}
\newcommand{  \Pb       }{\ifmmode \textrm{P}\beta \else P$\beta$\fi}
\newcommand{  \Bra      }{\ifmmode \textrm{Br}\alpha \else Br$\alpha$\fi}
\newcommand{  \Brg      }{\ifmmode \textrm{Br}\gamma \else Br$\gamma$\fi}
\newcommand{  \hii      }{\ifmmode \textrm{H}\,\textsc{ii} \else H\,\textsc{ii}\fi}
\newcommand{  \hei      }{\ifmmode \textrm{He}\,\textsc{i} \else He\,\textsc{i}\fi}
\newcommand{  \heii     }{\ifmmode \textrm{He}\,\textsc{ii} \else He\,\textsc{ii}\fi}
\newcommand{  \HeIIuv   }{\ifmmode \textrm{He}\,\textsc{ii}\,\lambda1640 \else He\,\textsc{ii}\,$\lambda1640$\fi}
\newcommand{  \HeIIop   }{\ifmmode \textrm{He}\,\textsc{ii}\,\lambda4686 \else He\,\textsc{ii}\,$\lambda4686$\fi}
\newcommand{  \CII	}{\ifmmode \left[\textrm{C}\,\textsc{ii}\right]\,\lambda157.74\,\mu\textrm{m} \else [C\,{\sc ii}]\ $\lambda157.74\,\mu\textrm{m}$\fi}
\newcommand{  \cii	}{\ifmmode \left[\textrm{C}\,\textsc{ii}\right] \else [C\,{\sc ii}]\fi}

\newcommand{  \ciii     }{\ifmmode \textrm{C}\,\textsc{iii}\right] \else C\,\textsc{iii}]\fi}
\newcommand{  \CIII     }{\ifmmode \textrm{C}\,\textsc{iii}\right]\,\lambda1909 \else C\,\textsc{iii}]\,$\lambda1909$\fi}
\newcommand{  \civ      }{\ifmmode \textrm{C}\,\textsc{iv}  \else C\,\textsc{iv}\fi}
\newcommand{  \CIV      }{\ifmmode \textrm{C}\,\textsc{iv}\,\lambda1549 \else C\,\textsc{iv}\,$\lambda1549$\fi}
\newcommand{  \NIIopt   }{\ifmmode \left[\textrm{N}\,\textsc{ii}\right]\,\lambda6584 \else [N\,\textsc{ii}]\,$\lambda6584$\fi}
\newcommand{  \nii      }{\ifmmode \left[\textrm{N}\,\textsc{ii}\right]  \else [N\,\textsc{ii}]\fi}
\newcommand{  \niii     }{\ifmmode \textrm{N}\,\textsc{iii} \else N\,\textsc{iii}\fi}
\newcommand{  \NIII     }{\ifmmode \textrm{N}\,\textsc{iii}\,\lambda4640 \else N\,\textsc{iii}\,$\lambda4640$\fi}
\newcommand{  \niv      }{\ifmmode \textrm{N}\,\textsc{iv}  \else N\,\textsc{iv}\fi}
\newcommand{  \NIVuv    }{\ifmmode \textrm{N}\,\textsc{iv}\,\lambda1486 \else N\,\textsc{iv}\,$\lambda1486$\fi}
\newcommand{  \nv       }{\ifmmode \textrm{N}\,\textsc{v}   \else N\,\textsc{v}\fi}
\newcommand{\oi}{\ifmmode \left[\textrm{O}\,\textsc{i}\right] \else [O\,{\sc i}]\fi}
\newcommand{\OI}{\ifmmode \left[\textrm{O}\,\textsc{i}\right]\,\lambda6300 \else [O\,{\sc i}]$\,\lambda6300$\fi}
\newcommand{\oii}{\ifmmode \left[\textrm{O}\,\textsc{ii}\right] \else [O\,{\sc ii}]\fi}
\newcommand{\OII}{\ifmmode \left[\textrm{O}\,\textsc{ii}\right]\,\lambda3727 \else [O\,{\sc ii}]\,$\lambda3727$\fi}
\newcommand{\oiii}{\ifmmode \textrm{O}\,\textsc{iii} \else O\,{\sc iii}\fi}
\newcommand{\OIII}{\ifmmode \left[\textrm{O}\,\textsc{iii}\right]\,\lambda5007 \else [O\,{\sc iii}]\,$\lambda5007$\fi}
\newcommand{  \OIIIbf   }{\ifmmode \textrm{O}\,\textsc{iii}\,\lambda3133 \else O\,\textsc{iii}\,$\lambda3133$\fi}
\newcommand{  \OIIIuv   }{\ifmmode \textrm{O}\,\textsc{iii}\,\lambda1663 \else O\,\textsc{iii}\,$\lambda1663$\fi}
\newcommand{  \oiv      }{\ifmmode \textrm{O}\,\textsc{iv}  \else O\,\textsc{iv}\fi}
\newcommand{  \OIVuv    }{\ifmmode \textrm{O}\,\textsc{iv}\,\lambda1402  \else O\,\textsc{iv}\,$\lambda1402$\fi}
\newcommand{  \OIVIR    }{\ifmmode \textrm{O}\,\textsc{iv}\,25.9\,\mu \textrm{m} \else O\,\textsc{iv}\,$25.9\,\mu$m\fi}
\newcommand{  \ovi      }{\ifmmode \textrm{O}\,\textsc{vi}   \else O\,\textsc{vi}\fi}
\newcommand{  \Ovi      }{\ifmmode \textrm{O}\,\textsc{vi}\,\lambda1035 \else O\,\textsc{vi}\,$\lambda1035$\fi}
\newcommand{  \nei      }{\ifmmode \textrm{Ne}\,\textsc{i}   \else Ne\,\textsc{i}\fi}
\newcommand{  \neii     }{\ifmmode \textrm{Ne}\,\textsc{ii}  \else Ne\,\textsc{ii}\fi}
\newcommand{  \NeiiIR   }{\ifmmode \textrm{Ne}\,\textsc{ii}\,12.8\,\mu \textrm{m} \else Ne\,\textsc{ii}\,$12.8\,\mu$m\fi}
\newcommand{  \neiii    }{\ifmmode \textrm{Ne}\,\textsc{iii} \else Ne\,\textsc{iii}\fi}
\newcommand{  \neiv     }{\ifmmode \textrm{Ne}\,\textsc{iv}  \else Ne\,\textsc{iv}\fi}
\newcommand{  \nev      }{\ifmmode \textrm{Ne}\,\textsc{v}   \else Ne\,\textsc{v}\fi}
\newcommand{  \NevIR    }{\ifmmode \textrm{Ne}\,\textsc{v}\,24.3\,\mu \textrm{m} \else Ne\,\textsc{v}\,$24.3\,\mu$m\fi}
\newcommand{  \nevi     }{\ifmmode \textrm{Ne}\,\textsc{vi}  \else Ne\,\textsc{vi}\fi}
\newcommand{  \mgi      }{\ifmmode \textrm{Mg}\,\textsc{i} \else Mg\,\textsc{i}\fi}
\newcommand{  \mgii     }{\ifmmode \textrm{Mg}\,\textsc{ii} \else Mg\,\textsc{ii}\fi}
\newcommand{  \MgII     }{\ifmmode \textrm{Mg}\,\textsc{ii}\,\lambda2798 \else Mg\,\textsc{ii}\,$\lambda2798$\fi}
\newcommand{  \sii      }{\ifmmode \textrm{S}\,\textsc{ii} \else S\,\textsc{ii}\fi}
\newcommand{  \siii     }{\ifmmode \textrm{S}\,\textsc{iii} \else S\,\textsc{iii}\fi}
\newcommand{  \siv      }{\ifmmode \textrm{S}\,\textsc{iv} \else S\,\textsc{iv}\fi}
\newcommand{  \sili     }{\ifmmode \textrm{Si}\,\textsc{i}   \else Si\,\textsc{i}\fi}
\newcommand{  \silii    }{\ifmmode \textrm{Si}\,\textsc{ii}  \else Si\,\textsc{ii}\fi}
\newcommand{  \Siliv    }{\ifmmode \textrm{Si}\,\textsc{iv}  \else Si\,\textsc{iv}\fi}
\newcommand{  \SilIVuv  }{\ifmmode \textrm{Si}\,\textsc{iv}\,\lambda1400  \else Si\,\textsc{iv}\,$\lambda1400$\fi}
\newcommand{  \AlIII   }{\ifmmode \textrm{Al}\,\textsc{iii}\,\lambda1857 \else Al\,\textsc{iii}\,$\lambda1857$\fi}
\newcommand{  \Aliii   }{\ifmmode \textrm{Al}\,\textsc{iii} \else Al\,\textsc{iii}\fi}
\newcommand{  \caii     }{\ifmmode \textrm{Ca}\,\textsc{ii} \else Ca\,\textsc{ii}\fi}
\newcommand{  \feii     }{\ifmmode \textrm{Fe}\,\textsc{ii} \else Fe\,\textsc{ii}\fi}
\newcommand{  \feiii    }{\ifmmode \textrm{Fe}\,\textsc{iii} \else Fe\,\textsc{iii}\fi}
\newcommand{  \Kalpha   }{\ifmmode \textrm{K}\alpha \else K$\alpha$\fi}

\newcommand{ \Lhb   }{\ifmmode L_{\hb} \else $L_{\hb}$\fi}
\newcommand{ \Lha   }{\ifmmode L_{\ha} \else $L_{\ha}$\fi}
\newcommand{ \fwhb  }{\ifmmode \textrm{FWHM}\left(\hb\right) \else  FWHM(\hb)\fi}
\newcommand{\sighb  }{\ifmmode \sigma\left(\hb\right) \else $\sigma\left(\hb\right)$\fi}
\newcommand{ \ewhb  }{\ifmmode \textrm{EW}\left(\hb\right) \else EW(\hb)\fi}
\newcommand{ \fwha  }{\ifmmode \textrm{FWHM}\left(\ha\right) \else FWHM(\ha)\fi}
\newcommand{ \ewha  }{\ifmmode \textrm{EW}\left(\ha\right) \else EW(\ha)\fi}
\newcommand{ \Lmg   }{\ifmmode L\left(\mgii\right) \else $L\left(\mgii\right)$\fi}
\newcommand{ \fwmg  }{\ifmmode \textrm{FWHM}\left(\mgii\right) \else FWHM(\mgii)\fi}
\newcommand{ \Lciv  }{\ifmmode L\left(\civ\right) \else $L\left(\civ\right)$\fi}
\newcommand{ \fwciv }{\ifmmode \textrm{FWHM}\left(\civ\right) \else FWHM(\civ)\fi}
\newcommand{ \fwhm  }{\ifmmode \textrm{FWHM} \else FWHM\fi} 
\newcommand{ \voff  }{\ifmmode v_\textrm{off} \else $v_\textrm{off}$\fi} 
\newcommand{ \vmax  }{\ifmmode v_\textrm{max} \else $v_\textrm{max}$\fi} 

\newcommand{ \mumg  }{\ifmmode \mu\left(\mgii\right) \else $\mu\left(\mgii\right)$\fi}
\newcommand{ \fmg   }{\ifmmode f\left(\mgii\right) \else $f\left(\mgii\right)$\fi}
\newcommand{ \muciv }{\ifmmode \mu\left(\civ\right) \else $\mu\left(\civ\right)$\fi}
\newcommand{ \fciv  }{\ifmmode f\left(\civ\right) \else $f\left(\civ\right)$\fi}
\newcommand{  \auvo     }{\ifmmode \alpha_{\nu,\textrm{UVO}} \else $\alpha_{\nu,\textrm{UVO}}$\fi}
\newcommand{  \aox      }{\ifmmode \alpha_{\,\textrm{O\textsc{x}}} \else $\alpha_{\,\textrm{O\textsc{x}}}$\fi}
\newcommand{  \Ledd     }{\ifmmode L_\textrm{Edd} \else $L_\textrm{Edd}$\fi}
\newcommand{  \lamLlam  }{\ifmmode \lambda L_{\lambda} \else $\lambda L_{\lambda}$\fi}
\newcommand{  \lLl      }{\ifmmode \lambda L_{\lambda} \else $\lambda L_{\lambda}$\fi}
\newcommand{  \nuLnu    }{\ifmmode \nu L_{\nu} \else $\nu L_{\nu}$\fi}
\newcommand{  \nLn      }{\ifmmode \nu L_{\nu} \else $\nu L_{\nu}$\fi}
\newcommand{  \Luv      }{\ifmmode L_{1450} \else $L_{1450}$\fi}
\newcommand{  \Lop      }{\ifmmode L_{5100} \else $L_{5100}$\fi}
\newcommand{  \lLop     }{\ifmmode \log\left(\Lop/\ergs\right) \else $\log\left(\Lop/\ergs\right)$\fi}
\newcommand{  \Lthree   }{\ifmmode L_{3000} \else $L_{3000}$\fi}
\newcommand{  \lLthree  }{\ifmmode \log\left(\Lthree/\ergs\right) \else $\log\left(\Lthree/\ergs\right)$\fi}
\newcommand{  \Lsix      }{\ifmmode L_{6200} \else $L_{6200}$\fi}
\newcommand{  \lLisx     }{\ifmmode \log\left(\Lop/\ergs\right) \else $\log\left(\Lop/\ergs\right)$\fi}
\newcommand{  \Lxray    }{\ifmmode L_\textrm{X} \else $L_\textrm{X}$\fi}

\newcommand{  \Lhard    }{\ifmmode L_\textrm{2-10} \else $L_\textrm{2-10}$\fi}
\newcommand{  \Lsoft    }{\ifmmode L_\textrm{0.5-2} \else $L_\textrm{0.5-2}$\fi}

\newcommand{\Fthree}{\ifmmode F_{3000} \else $F_{3000}$\fi}
\newcommand{\fuv}{\ifmmode f_{\lambda}\left(1450\textrm{\AA}\right) \else $f_{\lambda}\left(1450 \textrm{\AA}\right)$\fi}
\newcommand{\fthree}{\ifmmode f_{\lambda}\left(3000\textrm{\AA}\right) \else $f_{\lambda}\left(3000\textrm{\AA}\right)$\fi}
\newcommand{\fH}{\ifmmode f_{\lambda}\left(1.65\micron\right) \else
$f_{\lambda}\left(1.65\micron\right)$\fi}

\newcommand{\fbol}{\ifmmode f_\textrm{bol} \else $f_\textrm{bol}$\fi}
\newcommand{\fbolwv}{\ifmmode f_\textrm{bol}\left(\lambda\right) \else $f_\textrm{bol}\left(\lambda\right)$\fi}
\newcommand{\fbolopt}{\ifmmode f_\textrm{bol}\left(5100\textrm{\AA}\right) \else $f_\textrm{bol}\left(5100\textrm{\AA}\right)$\fi}
\newcommand{\fbolthree}{\ifmmode f_\textrm{bol}\left(3000\textrm{\AA}\right) \else $f_\textrm{bol}\left(3000\textrm{\AA}\right)$\fi}
\newcommand{\fboluv}{\ifmmode f_\textrm{bol}\left(1450\textrm{\AA}\right) \else $f_\textrm{bol}\left(1450\textrm{\AA}\right)$\fi}

\newcommand{\fbolbat}{\ifmmode f_\textrm{bol}\left(14-150\,\kev\right) \else $f_\textrm{bol}\left(14-150\,\kev\right)$\fi}
\newcommand{\fbolhard}{\ifmmode f_\textrm{bol}\left(2-10\,\kev\right) \else $f_\textrm{bol}\left(2-10\,\kev\right)$\fi}

\newcommand{\fobs}{\ifmmode f_\textrm{obs} \else $f_\textrm{obs}$\fi}

\newcommand{\mbh}{\ifmmode M_{\rm BH} \else $M_{\rm BH}$\fi}
\newcommand{  \lmbh     }{\ifmmode \log\left(\mbh/\Msun\right) \else $\log\left(\mbh/\Msun\right)$\fi} 
\newcommand{  \lledd    }{\ifmmode L/L_\textrm{Edd} \else $L/L_\textrm{Edd}$\fi}
\newcommand{  \mmedd    }{\ifmmode \dot{m}/\dot{m}_\textrm{\,Edd} \else $\dot{m}/\dot{m}_\textrm{\,Edd}$\fi}
\newcommand{  \Lbol     }{\ifmmode L_\textrm{bol} \else $L_\textrm{bol}$\fi}
\newcommand{  \lbol     }{\ifmmode L_\textrm{bol} \else $L_\textrm{bol}$\fi}
\newcommand{  \lLbol    }{\ifmmode \log\left(\Lbol/\ergs\right) \else $\log\left(\Lbol/\ergs\right)$\fi} 
\newcommand{  \Lagn     }{\ifmmode L_\textrm{AGN} \else $L_\textrm{AGN}$\fi}
\newcommand{  \lagn     }{\ifmmode L_\textrm{AGN} \else $L_\textrm{AGN}$\fi}

\newcommand{  \tgrow     }{\ifmmode t_\textrm{growth} \else $t_\textrm{growth}$\fi}
\newcommand{  \tAD     }{\ifmmode t_\textrm{acc} \else $t_\textrm{acc}$\fi}
\newcommand{  \tacc    }{\ifmmode t_\textrm{acc} \else $t_\textrm{acc}$\fi}
\newcommand{  \tUni      }{\ifmmode t_\textrm{Universe} \else $t_\textrm{Universe}$\fi}

\newcommand{  \Mdotin	}{\ifmmode \dot{M}_\textrm{infall} \else $\dot{M}_\textrm{infall}$\fi}
\newcommand{  \Mdotbh	}{\ifmmode \dot{M}_\textrm{BH} \else $\dot{M}_\textrm{BH}$\fi}
\newcommand{  \Mdotad	}{\ifmmode \dot{M}_\textrm{AD} \else $\dot{M}_\textrm{AD}$\fi}
\newcommand{  \Mdotacc	}{\ifmmode \dot{M}_\textrm{acc} \else $\dot{M}_\textrm{acc}$\fi}
\newcommand{  \Mdotthin	}{\ifmmode \dot{M}_\textrm{thin} \else $\dot{M}_\textrm{thin}$\fi}
\newcommand{  \Mdotdisk	}{\ifmmode \dot{M}_\textrm{disk} \else $\dot{M}_\textrm{disk}$\fi}

\newcommand{  \Mindot	}{\ifmmode \dot{M}_\textrm{infall} \else $\dot{M}_\textrm{infall}$\fi}
\newcommand{  \Mbhdot	}{\ifmmode \dot{M}_\textrm{BH} \else $\dot{M}_\textrm{BH}$\fi}
\newcommand{  \Maddot	}{\ifmmode \dot{M}_\textrm{AD} \else $\dot{M}_\textrm{AD}$\fi}
\newcommand{  \Maccdot	}{\ifmmode \dot{M}_\textrm{acc} \else $\dot{M}_\textrm{acc}$\fi}
\newcommand{  \Mthdot	}{\ifmmode \dot{M}_\textrm{thin} \else $\dot{M}_\textrm{thin}$\fi}
\newcommand{  \Mdsdot	}{\ifmmode \dot{M}_\textrm{disk} \else $\dot{M}_\textrm{disk}$\fi}

\newcommand{  \as	}{\ifmmode a_\textrm{*} \else $a_\textrm{*}$\fi}
\newcommand{  \avec	}{\ifmmode \vec{a}_\textrm{*} \else $\vec{a}_\textrm{*}$\fi}
\newcommand{  \re	}{\ifmmode \eta      	 \else $\eta$\fi}
\newcommand{  \RISCO	}{\ifmmode R_\textrm{ISCO}  \else $R_\textrm{ISCO}$\fi}

\newcommand{  \mseed    }{\ifmmode M_\textrm{seed} \else $M_\textrm{seed}$\fi}
\newcommand{  \mbul     }{\ifmmode M_\textrm{bulge} \else $M_\textrm{bulge}$\fi} 
\newcommand{  \mstar    }{\ifmmode M_{*} \else $M_{*}$\fi} 
\newcommand{  \rstar    }{\ifmmode R_{*} \else $R_{*}$\fi} 
\newcommand{  \mgal     }{\ifmmode M_{*} \else $M_{*}$\fi} 
\newcommand{  \mhost    }{\ifmmode M_\textrm{host} \else $M_\textrm{host}$\fi}
\newcommand{  \mmsmall  }{\ifmmode M_\textrm{BH}/M_{*} \else $M_\textrm{BH}/M_{*}$\fi}
\newcommand{  \mmlarge  }{\ifmmode M_{*}/M_\textrm{BH} \else $M_{*}/M_\textrm{BH}$\fi}

\newcommand{  \mmdotlarge}{\ifmmode \dot{M}_*/\Mbhdot \else $\dot{M}_*/\Mbhdot$\fi}
\newcommand{  \mmdotsmall}{\ifmmode \Mbhdot/\dot{M}_* \else $\Mbhdot/\dot{M}_*$\fi}

\newcommand{  \mmwp     }{\ifmmode \left(M_{*}/M_\textrm{BH}\right) \else $\left(M_{*}/M_\textrm{BH}\right)$\fi}
\newcommand{  \ml       }{\ifmmode M_{*}/L_{*} \else $M_{*}/L_{*}$\fi}
\newcommand{  \mlwp     }{\ifmmode \left(M_{*}/L\right) \else $\left(M_{*}/L\right)$\fi}
\newcommand{  \mlk      }{\ifmmode \left(M_{*}/L_{K}\right) \else $\left(M_{*}/L_{K}\right)$\fi}
\newcommand{  \sigs     }{\ifmmode \sigma_{*} \else $\sigma_{*}$\fi}
\newcommand{  \Reff     }{\ifmmode R_\textrm{e} \else $R_\textrm{e}$\fi}
\newcommand{  \Rvir     }{\ifmmode R_\textrm{vir} \else $R_\textrm{vir}$\fi}
\newcommand{  \Rtwo     }{\ifmmode R_{200} \else $R_{200}$\fi}
\newcommand{  \Rfive    }{\ifmmode R_{500} \else $R_{500}$\fi}
\newcommand{  \Rgrp     }{\ifmmode R_\textrm{grp} \else $R_\textrm{grp}$\fi}
\newcommand{  \nser     }{\ifmmode n_\textrm{s} \else $n_\textrm{s}$\fi}
\newcommand{  \LSF      }{\ifmmode L_\textrm{SF}  \else $L_\textrm{SF}$\fi}
\newcommand{  \LFIR     }{\ifmmode L_\textrm{FIR} \else $L_\textrm{FIR}$\fi}
\newcommand{  \Lfir     }{\ifmmode L_\textrm{FIR} \else $L_\textrm{FIR}$\fi}
\newcommand{  \LTIR     }{\ifmmode L_\textrm{TIR} \else $L_\textrm{TIR}$\fi}
\newcommand{  \Ltir     }{\ifmmode L_\textrm{TIR} \else $L_\textrm{TIR}$\fi}

\newcommand{  \mdyn     }{\ifmmode M_\textrm{dyn} \else $M_\textrm{dyn}$\fi} 
\newcommand{  \mgas     }{\ifmmode M_\textrm{gas} \else $M_\textrm{gas}$\fi} 
\newcommand{  \mh       }{\ifmmode M_\textrm{h} \else $M_\textrm{h}$\fi}
\newcommand{  \mhalo    }{\ifmmode M_\textrm{halo} \else $M_\textrm{halo}$\fi}
\newcommand{  \sfr      }{\ifmmode \textrm{SFR} \else SFR\fi}

\newcommand{ \Lcii     }{\ifmmode L_{\cii} \else $L_{\cii}$\fi}
\newcommand{ \fwcii  }{\ifmmode \textrm{FWHM}\cii \else FWHM\cii\fi}

\newcommand{\Rin}{\ifmmode R_\textrm{in} \else $R_\textrm{in}$\fi}
\newcommand{\Rout}{\ifmmode R_\textrm{out} \else $R_\textrm{out}$\fi}
\newcommand{\RBLR}{\ifmmode R_\textrm{BLR} \else $R_\textrm{BLR}$\fi}
\newcommand{\RNLR}{\ifmmode R_\textrm{NLR} \else $R_\textrm{NLR}$\fi}

\newcommand{  \swift   }  {{\it Swift}}

\newcommand{\bj}{\ifmmode b_\textrm{J} \else $b_\textrm{J}$\fi}

\newcommand{\iab}{\ifmmode i_\textrm{AB} \else $i_\textrm{AB}$\fi}

\newcommand{\jab}{\ifmmode J_\textrm{AB} \else $J_\textrm{AB}$\fi}
\newcommand{\hab}{\ifmmode H_\textrm{AB} \else $H_\textrm{AB}$\fi}
\newcommand{\kab}{\ifmmode K_\textrm{AB} \else $K_\textrm{AB}$\fi}

\newcommand{\jveg}{\ifmmode J_\textrm{Vega} \else $J_\textrm{Vega}$\fi}
\newcommand{\hveg}{\ifmmode H_\textrm{Vega} \else $H_\textrm{Vega}$\fi}
\newcommand{\kveg}{\ifmmode K_\textrm{Vega} \else $K_\textrm{Vega}$\fi}

\newcommand{  \Chisq    }{\ifmmode \chi^{2} \else $\chi^{2}$}
\newcommand{  \nelec    }{\ifmmode n_{e} \else $n_{e}$\fi}     % electron density
\newcommand{  \nh       }{\ifmmode n_{H} \else $n_{H}$\fi}     % hydrogen density
\newcommand{  \Ncol     }{\ifmmode N_\textrm{col} \else $N_\textrm{col}$\fi} % column density
\newcommand{  \NH       }{\ifmmode N_\textrm{H} \else $N_\textrm{H}$\fi}     % column density

\def\sun{\hbox{$\odot$}}

\def\arcsec{\hbox{$^{\prime\prime}$}}
\def\ion#1#2{#1$\;${\small\textrm{\@Roman{#2}}}\relax}